# Multi-Material Topology Optimization for Targeted Structural Dynamic Vibration Attenuation

Bir Bahadur Chaudhary [a], Bidhan Pandey [a], Bin Ouyang [b], Shiguang Deng [a*]
[a] Department of Mechanical Engineering, University of Kansas, Lawrence, KS, US
[b] Department of Chemical and Biomolecular Engineering, Vanderbilt University, Nashville, TN, USA

## Abstract

Resonance-induced excessive vibrations can degrade performance, reduce reliability, and shorten the service life of engineering structures. To reduce the risk of resonance, structural natural frequencies are generally engineered to be shifted away from excitation frequencies. To achieve these frequency shifts, topology optimization provides a systematic approach to tailor structural stiffness and mass as an effective inverse design approach. Multi-material topology optimization exploits the high-dimensional design space of spatially varying material constituents by simultaneously optimizing material selection and spatial distribution over the design domain, offering great flexibility in controlling natural frequencies. This study proposes a multi-material topology optimization that attenuates excessive structural dynamic vibrations in free-vibration scenarios via a novel multivariate weight-based interpolation method. Additionally, to address non-smoothness due to mode switching amid optimization, we employ a bound formulation with an auxiliary scalar design variable to replace the max-min eigenvalue objective with constraints on individual eigenvalues. Furthermore, we study and compare three optimization formulations for multi-material design to attenuate vibrations, including maximizing fundamental eigenfrequency, maximizing the gap between adjacent eigenfrequencies, and excluding frequencies from targeted frequency bands, which are smoothly approximated by a proposed rational polynomial formulation. Numerical examples are presented to demonstrate the effectiveness and applicability of our proposed method.

## 1. Introduction

Dynamic vibrations significantly affect the performance, integrity, and operational lifespan of engineered structures, where unmitigated responses can cause severe structural degradation and failure [1]. Developing effective vibration attenuation strategies has become a central objective in structural design and optimization [2]. Engineering structures, such as beams, plates, and shells, which are widely used in aircraft structures, spacecraft structures, bridges, and machinery, are inevitably subjected to dynamic excitation during operation [3]. The resulting vibration response is strongly influenced by the relationship between the external excitation frequencies and the natural frequencies, and corresponding mode shapes. When an excitation frequency coincides with a natural frequency and couples with corresponding mode shapes, resonance produces large-amplitude oscillations leading to excessive deformation [4]. Resonance-induced vibration relies on the distribution of natural frequencies: when the excitation is close to a single natural frequency, the corresponding mode generally dominates, whereas closely spaced natural frequencies or repeated eigenvalues cause multiple modes to participate simultaneously through modal superposition, resulting in complex dynamic behaviors [5], [6].

Vibration reduction can improve fatigue resistance, lower acoustic emission, enhance reliability, and prolong structure life; yet conventional mitigation strategies generally carry limitations. Passive mitigation approaches, such as structural damping enhancement and vibration isolation techniques, are widely adopted; such approaches, however, need to add additional masses and are generally effective over limited frequency ranges [7]. Active control schemes offer broadband performance, but require sensors, actuators, and energy supplies, thereby increasing complexity and cost [8]. An alternative is to design the structure such that its natural frequencies remain outside the operating frequency range to prevent resonance [9]. This approach is generally pursued by maximizing the fundamental eigenfrequency [10].

Topology optimization (TO) enables the systematic tailoring of dynamic characteristics through the distribution of material constituents and voids within a prescribed domain [11], [12], [13], [14]. Popular methods include the homogenization method [15], density-based formulation [16], [17], evolutionary approach [18], level-set [19], [20], moving morphable components (MMCs) [21], geometric projection method [22], and genetic algorithm [23], [24]. While these methods generally differ in design representation and optimization updates, their objective and constraints can be adapted to account for structural vibration characteristics [25]. The homogenization method was applied to maximize the fundamental natural frequency by optimizing the distribution of reinforcing material [26]. It was subsequently extended from individual-frequency improvement to spectral tailoring through the optimization of multiple eigenfrequencies [27]. Design sensitivity and optimality were established for repeated eigenvalues to address the difficulty of non-differentiability [28], and artificial localized modes in low-density regions in density-based methods [29].

Recent studies have extended eigenfrequency-based topology optimization to targeted vibration suppression [30] and frequency-dependent material properties [4]. Harmonic-response optimization incorporated a modal assurance criterion (MAC)-based mode-tracking strategy, where the excitation frequency is progressively increased toward the prescribed value amid optimization, penalizing structural optimization to avoid resonance [31]. While this approach targets the response at a prescribed excitation frequency, resonance avoidance over an operating frequency band requires the control of natural frequencies whose mode orders may not be known in advance. To address this issue, a frequency-band constraint was proposed to prevent natural frequencies from entering prescribed bands via Heaviside functions [32].

*Corresponding author: sdeng@ku.edu

Compared to single-material topology optimization, multi-material topology optimization (MMTO) expands design space by optimizing structural performance through material constituents and spatial distribution. Early developments included peak-function interpolation to represent materials without increasing the number of design variables [33], rational approximation of material properties (RAMP) to accommodate an arbitrary number of isotropic and anisotropic candidate materials [34], ordered SIMP to represent multiple materials using a single elemental density variable [35], and a univariate mapping-based interpolation scheme to reduce intermediate material densities and prevent undesired layers between material phases by integrating Heaviside projections with multi-variable filtering [36]. Additionally, to enforce discrete material selection per finite element, multiphase level-set methods distinguish non-overlapping material regions [37], and phase-field formulations regularize through energy-based models [38], [39].

MMTO has been investigated for vibration attenuation, where defining proper design objectives relies on the relationship between the structural spectrum and operating frequencies. In a pioneering multi-material beam design [40], two approaches were taken: maximizing the fundamental eigenfrequency to raise the lowest resonance frequency above an operating range, and maximizing the separation between adjacent eigenfrequencies to create an interval between resonance frequencies. Dynamic MMTO has also been extended to multiple materials with frequency constraints using an alternating active phase algorithm (AAPA) interpolation scheme equipped with penalized interpolation for stiffness and linear interpolation for mass to address pseudo modes and mode switching amid optimization [41]. Simultaneous interpolation of stiffness and mass is particularly important in eigenvalue-based MMTO, because both quantities directly influence the natural frequencies and associated mode shapes. Many existing studies have considered different objectives, such as maximizing fundamental frequency, separating adjacent eigenfrequencies, and avoiding resonance in prescribed operating ranges. However, these objectives generally suffer from challenges, such as mode switching in eigenvalue maximization and effective determination of natural frequencies in a prescribed interval.

This study develops a multi-material topology optimization formulation to reduce resonance-induced excessive vibration with the following main contributions. First, we propose a novel multivariate weight-based interpolation scheme for multi-materials for stiffness and mass interpolation in eigenvalue-based MMTO. Second, we study and compare three distinct MMTO formulations based on the proposed interpolation scheme, including fundamental frequency maximization, adjacent-frequency gap maximization, and targeted frequency band generation. Third, we propose a smooth frequency band constraint based on rational polynomial approximation to address non-differentiability of frequency-band constraints.

We organize the remainder of this paper as follows. In ***Section 2***, we introduce the mathematical formulation of structural vibration. In ***Section 3***, we propose a multi-material interpolation scheme for stiffness and mass matrices. ***Section 4*** presents the topology optimization formulations for eigenfrequency optimization. In ***Section 5***, we derive design sensitivities. We describe the numerical implementation in ***Section 6*** and demonstrate the proposed approach through numerical examples in ***Section 7***. Finally, we summarize contributions in ***Section 8***.

## 2. Mathematical model of structural vibration

In this section, we model structural vibration by the finite element method (FEM) under the assumptions of linear elasticity, small deformations, and undamped free vibration. Accordingly, we neglect all external forces and damping effects, so that the dynamic response is governed by the mass and stiffness distribution of the structure. This reduces the problem to a generalized eigenvalue problem, from which we extract the natural frequencies and the corresponding mode shapes. In the following sections, we present the governing equations of structural free vibration in ***Section 2.1***, followed by the density filtering and Heaviside projection approaches in ***Section 2.2*** to prevent checkerboard patterns, reduce mesh dependency, and promote distinct material distributions during optimization

### 2.1. Governing equation for structural vibration

Undamped free vibration can be described by the second-order equation of motion given in ***Equation (1):***

$$\boldsymbol{M}\ddot{\boldsymbol{u}} + \boldsymbol{K}\boldsymbol{u} = \boldsymbol{0}, \tag{1}$$

where $\boldsymbol{M}$ and $\boldsymbol{K}$ denote the global mass and stiffness matrices, respectively, and $\boldsymbol{u}$ is the vector of nodal displacements. Since no external excitation or dissipative term appears in ***Equation (1),*** the motion is sustained entirely by the exchange between kinetic and strain energy, and the resulting oscillation is harmonic in time. ***Equation (1)*** therefore constitutes a second-order differential equation with constant coefficients, whose general solution can be expressed in the separable form $\boldsymbol{u} = \boldsymbol{v}\, e^{i\omega t}$ , where $\boldsymbol{v}$ is an eigenvector describing the spatial distribution of the motion and $\omega$ is the associated eigenfrequency. Substituting this solution into ***Equation (1)***, followed by elimination of the non-vanishing factor $e^{i\omega t}$, we remove the time dependence and reduce the problem to the generalized eigenvalue problem of ***Equation (2)***:

$$(\boldsymbol{K} - \omega^2\boldsymbol{M})\boldsymbol{v}_j = \boldsymbol{0};\ \ j = 1, 2, \dots, n_d, \tag{2}$$

where $n_d$ is the degree of freedom in the discretized model. A non-trivial solution exists only when the coefficient matrix becomes non-singular, which yields $n_d$ eigenpairs $(\omega_j^2, \boldsymbol{v}_j)$ corresponding to the natural frequencies and mode shapes of the vibration. The eigenvectors are normalized with respect to the mass matrix according to ***Equation (3)***. This orthonormality ensures that the eigenvectors form a linearly independent basis.

$$\boldsymbol{v}_i^T \boldsymbol{M} \boldsymbol{v}_j = \begin{cases} 0 & i \neq j \\ 1 & i = j. \end{cases} \tag{3}$$

### 2.2. Design variable filtering and projection

In this work, to prevent checkerboard patterns and mesh dependence, density filtering is applied to each design variable within a prescribed filter radius. A smoothed Heaviside projection is then applied to the filtered density design variables to reduce intermediate values and obtain a discrete material type, with the projection steepness gradually increased through a continuation scheme.

#### 2.2.1. Density filter

Element-wise density formulation is prone to checkerboard patterns and mesh dependence. To suppress both, a density filter [42], [43] is applied independently to each of the two design fields, giving the filtered densities ($\tilde{x}_e = [\tilde{x}_e^{(1)}, \tilde{x}_e^{(2)}]$) in ***Equation (4)***:

$$\tilde{x}_e^{(f)} = \frac{\sum_{j \in \mathcal{N}_e} \boldsymbol{H}_{ej} x_j^{(f)}}{\sum_{j \in \mathcal{N}_e} \boldsymbol{H}_{ej}}, \quad f = 1,2. \tag{4}$$

Here, $N_e$ denotes the set of elements whose centroids lie within the filter radius $r_{min}$ of the element $e$, and the linear weighting factor $H_{ej} = \max\left(0, r_{min} - \| r_e - r_j \|\right)$ decays with the distance between the centroids $r_e$ and $r_j$, so that neighboring elements at closer distances contribute more to filtered density.

#### 2.2.2. Heaviside projection

Although filtering suppresses numerical instabilities, it introduces intermediate-density regions near structural boundaries. As these regions lack clear physical meaning and may complicate manufacturing, a smoothed Heaviside projection [44], given in ***Equation (5)***, is applied to each filtered field to obtain the projected physical densities $\phi_e = [\phi_e^{(1)}, \phi_e^{(2)}]$.

$$\phi_e^{(f)} = \frac{\tanh(\beta_f \eta) + \tanh(\beta_f(\tilde{x}_e^{(f)} - \eta))}{\tanh(\beta_f \eta) + \tanh(\beta_f(1 - \eta))}, \tag{5}$$

where $\eta$ is the projection threshold and $\beta_f$ controls the sharpness of the projection. For small $\beta_f$, the mapping is nearly linear, whereas for large $\beta_f$, it approaches an ideal step function at $\eta$, driving densities towards 0 and 1. Imposing a large $\beta_f$ makes the optimization highly non-linear and causes divergence; $\beta_f$ is therefore increased gradually during the optimization following the continuation rule of [45], [46], [47], explained in ***Section 6***.

Since the objective and constraints are evaluated in terms of the projected densities, the sensitivities must be transferred back to the design variables through the chain rule. The required derivative of the projection is obtained by differentiating ***Equation (5)*** and is given in ***Equation (6)***.

$$\frac{\partial \phi_e^{(f)}}{\partial \tilde{x}_e^{(f)}} = \frac{\beta_f\left(1 - \tanh^2(\beta_f(\tilde{x}_e^{(f)} - \eta))\right)}{\tanh(\beta_f \eta) + \tanh(\beta_f(1 - \eta))}. \tag{6}$$

## 3. Multivariate weight-based multi-material interpolation

We discretize the design domain into $n_e = n_{elx} \times n_{ely}$ finite elements, where $n_{elx}$ and $n_{ely}$ denote the numbers of elements in the horizontal and vertical directions in a 2D design domain. We propose a novel formulation, i.e., a multivariate weight-based interpolation scheme to describe the distribution of candidate materials. For $n_{MAT}$ candidate materials, the scheme requires $n_{MAT} - 1$ design fields, and consequently, $n_{MAT} - 1$ design variables are assigned to each finite element. For example, for a three-material structure, it yields two design fields, so that the design variable of the element $e$ becomes $x_e = [x_e^{(1)}, x_e^{(2)}]$, with $x_e^{(1)}, x_e^{(2)} \in [0,1]$.

$$\mathrm{M}(x_e) = \begin{cases} \mathrm{M1}, & x_e^{(1)} < x_{e,th} \text{ and } x_e^{(2)} \in [0,1], \\ \mathrm{M2}, & x_e^{(1)} \geq x_{e,th} \text{ and } x_e^{(2)} < x_{e,th}, \\ \mathrm{M3}, & x_e^{(1)} \geq x_{e,th} \text{ and } x_e^{(2)} \geq x_{e,th}. \end{cases} \tag{7}$$

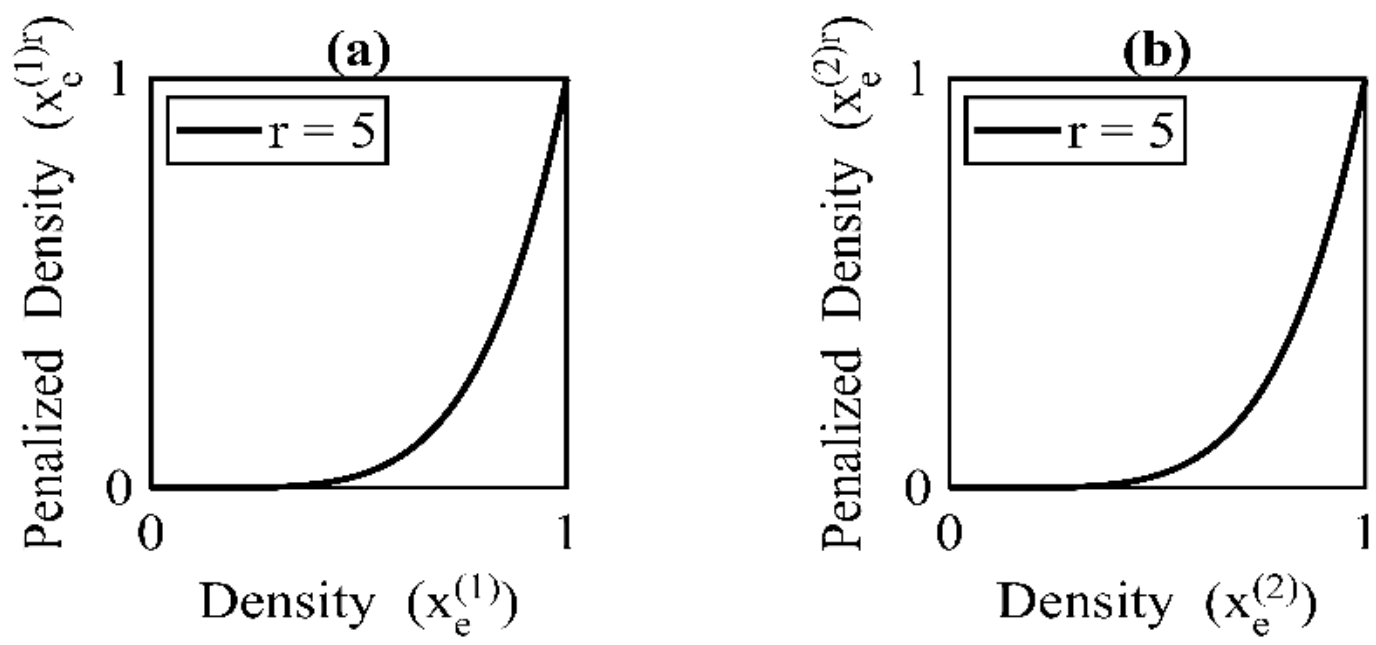

**Figure 1 Illustration of the penalization in the proposed multivariate weight-based interpolation: (a)** the first design variable in a three-material design, which selects between the first material and the remaining two materials, and **(b)** the second design variable, which selects between the second and third materials.

During optimization, design variables remain continuous, so the type of the material occupying each element is identified by comparing the design field against a threshold value of $x_{e,th} = 0.5$ as expressed in ***Equation (11)***. Values below the threshold are interpreted as the preceding candidate material and values above it as the succeeding one. The threshold is applied only to extract and report the material distribution, and does not enter the optimization; therefore, no non-differentiability is induced. For example, in an optimized three-material design, the interpolation scheme assigns the first type of candidate material (M1) per element when $x_e^{(1)} = 0$, the second material (M2) when $x_e^{(1)} = 1$ and $x_e^{(2)} = 0$, and the third material (M3) when $x_e^{(1)} = 1$ and $x_e^{(2)} = 1$, as expressed in ***Equation (11)***. ***Figure 1*** illustrates this penalization scheme for the three-material example. Specifically, ***Figure 1(a)*** shows the interpolation between the first and second materials, while ***Figure 1(b)*** shows the interpolation between the second and third materials. The selection of either M2 or M3 requires the corresponding condition in ***Figure 1(a)*** to be satisfied first, after which the second design field distinguishes between these two materials. This hierarchical procedure is extended to a system containing user-defined $n_{MAT}$ materials by successively applying the interpolation scheme to the candidate materials.

### 3.1. Weights for multi-material interpolation scheme

During optimization, the design variables are firstly filtered using ***Equation (4)*** and subsequently projected using ***Equation (5)***. The projected fields are then converted into material weights, which define the contribution of each candidate material to the effective element properties. For a mesh of $n_e$ elements and $n_{Mat}$ candidate materials, the unnormalized weight of the first material follows directly from the first projected field, as shown in ***Equation (8)***.

$$\boldsymbol{w}_1^{un} = \boldsymbol{\phi}_1^r. \tag{8}$$

The unnormalized weights for intermediate materials ($2 \leq m \leq n_{Mat} - 1$) are designed as in ***Equation (9)***:

$$\boldsymbol{w}_m^{un} = \left( \prod_{j=1}^{m-1} (1 - \boldsymbol{\phi}_j^r) \right) \boldsymbol{\phi}_m^r . \tag{9}$$

The unnormalized weight for the last material takes the remaining fraction, which is given by ***Equation (10)***:

$$\boldsymbol{w}_{n_{Mat}}^{un} = \prod_{j=1}^{n_{Mat}-1} (1 - \boldsymbol{\phi}_j^r) . \tag{10}$$

The weights are normalized according to ***Equation (11)*** to enforce the partition-of-unity condition. We note that this normalization guarantees that the material fractions within an element are mutually exclusive and complementary: an increase in the fraction of one material is necessarily accompanied by a corresponding decrease in the fractions of the remaining materials. Consequently, the interpolated element properties always remain bounded by those of the candidate materials.

$$\boldsymbol{w}_m = \frac{\boldsymbol{w}_m^{un}}{\sum_{m=1}^{n_{Mat}} \boldsymbol{w}_m^{un}} \quad \text{and} \quad \sum_{m=1}^{n_{Mat}} \boldsymbol{w}_m = 1. \tag{11}$$

### 3.2. Finite element formulation for structural vibration

We discretize the design domain using four-node bilinear quadrilateral elements, each having two translational degrees of freedom per node. Since the out-of-plane thickness of the domain is small relative to its in-plane dimensions, we assume a plane stress condition, so that the elastic modulus matrix of the $m^{th}$ candidate material is described by ***Equation (12)*** under the plane stress assumption:

$$\boldsymbol{D_m} = \frac{E_m}{1 - {\nu_m}^2} \begin{bmatrix} 1 & \nu_m & 0 \\ \nu_m & 1 & 0 \\ 0 & 0 & \frac{1 - \nu_m}{2} \end{bmatrix}, \tag{12}$$

where $E_m$ and $\nu_m$ denote the Young's modulus and Poisson's ratio of the material $m$, respectively. The corresponding element stiffness and mass matrices are obtained from the standard Galerkin formulation in ***Equation (13)***:

$$\boldsymbol{K_e}^{(m)} = \int_{\Omega_e} \boldsymbol{B}^T \boldsymbol{D_m} \boldsymbol{B} \, d\Omega; \;\; \boldsymbol{M_e}^{(m)} = \int_{\Omega_e} \rho_m \boldsymbol{N}^T \boldsymbol{N} \, d\Omega, \tag{13}$$

in which $\boldsymbol{B}$ is the strain-displacement matrix derived from the shape function matrix $\boldsymbol{N}$, $\rho_m$ is the mass density of the material $m$, and the integrations are carried out numerically over the element domain $\Omega_e$ using Gauss quadrature.

We assume all candidate materials share the same Poisson's ratio, allowing the material properties to be decoupled from the element geometry, so that the contribution of each element to the structural response can be expressed through an effective modulus of elasticity and an effective density, as stated in ***Equation (14)***:

$$E_e^{eff} \sum_{m=1}^{n_{Mat}} E_m \, \boldsymbol{w}_{e,m} \;\; ; \;\; \rho_e^{eff} = \sum_{m=1}^{n_{Mat}} \rho_m \, \boldsymbol{w}_{e,m}, \tag{14}$$

where $\boldsymbol{w}_{e,m}$ is the interpolation weight of material $m$ in element $e$, and $n_{Mat}$ is the number of candidate materials. These weights govern the transition between the constituent materials, thereby determining the amount of each material present in each element.

Consequently, global stiffness and mass matrices are assembled from the elemental contributions scaled by these effective properties in ***Equation (15)***:

$$\boldsymbol{K} = \sum_{e=1}^{n_e} E_e^{eff} \, \boldsymbol{K_e} \, ; \;\; \boldsymbol{M} = \sum_{e=1}^{n_e} \rho_e^{eff} \, \boldsymbol{M_e}, \tag{15}$$

where $\boldsymbol{K_e}$ is the elemental stiffness matrix evaluated for unit modulus of elasticity, $\boldsymbol{M_e}$ is the elemental mass matrix evaluated for unit material density, and $n_e$ is the total number of elements. Both matrices are computed once at the beginning of the analysis and reused in every iteration, which appreciably reduces the computational effort of the optimization.

With the global matrices assembled and the prescribed boundary conditions imposed, the generalized eigenvalue problem of ***Equation (2)*** is solved to obtain the eigenvalues along with associated eigenvectors. The eigenvectors are subsequently mass-normalized by ***Equation (3)***, providing a consistent basis required for sensitivity analysis in ***Section 5***.

### 3.3. Frequency band constraint by rational polynomial function

In many practical engineering applications, it is not sufficient merely to maximize the fundamental frequency or to widen the gap between adjacent eigenmodes. What ultimately governs the dynamic serviceability of a component is whether any of its natural frequencies coincides with the excitation frequencies generated during operation. A design possessing a high fundamental frequency may still be undesirable, if a higher mode happens to lie within the operating range. For this reason, the design requires that no natural frequency of the structure is permitted to fall inside prescribed working frequency ranges, so that resonance and the associated amplification of vibration amplitudes are avoided.

Let prohibited bands be defined by the lower and upper frequency limits ($\omega_{low}^{(b)}$ and $\omega_{upp}^{(b)}$) and indexed by $b \in \{1, \dots, n_M\}$, where $n_M$ is the number of bands. It is convenient to describe each band by its center and half-width by ***Equation (16)***:

$$\omega_c^{(b)} = \frac{\omega_{upp}^{(b)} + \omega_{low}^{(b)}}{2}, \;\; \omega_d^{(b)} = \frac{\omega_{upp}^{(b)} - \omega_{low}^{(b)}}{2}, \tag{16}$$

where this pair of quantities maps the band onto the normalized coordinate $\left(\omega - \omega_c^{(b)}\right)/\omega_d^{(b)}$, which equals zero at the band center and $\pm 1$ at the band edges. The exclusion requirement can be written compactly as $\omega_j \notin [\omega_{low}^{(b)}, \omega_{upp}^{(b)}]$ for every eigenfrequency $\omega_j$ by the Heaviside function in ***Equation (17)***:

$$H(\omega_j) = \begin{cases} 1, & \omega_j \in [\omega_{low}^{(b)}, \omega_{upp}^{(b)}] \\ 0, & \omega_j \notin [\omega_{low}^{(b)}, \omega_{upp}^{(b)}] \end{cases}, \tag{17}$$

which returns unity whenever a frequency violates the exclusion condition and zero otherwise. However, it cannot be used directly in a gradient-based optimization: the function is discontinuous at the band edges, its derivative vanishes everywhere else, and therefore, it supplies little indication about how far a given eigenfrequency lies from the prohibited region. For example, a design whose eigenfrequency sits just inside the band and another design whose eigenfrequency lies at the band center are indistinguishable to the optimizer, and therefore, no descent direction can be constructed.

To overcome this limitation, we replace the Heaviside function with a smooth and differentiable approximation: a rational polynomial function (RPF) defined in ***Equation (18)***:

$$RPF(\omega) = \frac{1}{1 + \left(\frac{\omega - \omega_c}{\omega_d}\right)^{2\beta_{FBC}}}, \tag{18}$$

in which $\beta_{FBC}$ is a positive integer that controls transition sharpness, as illustrated in ***Figure 2***. At the band center the normalized coordinate vanishes, and the function attains its maximum value of unity; at either band edge the normalized coordinate equals $\pm 1$, so that the denominator becomes exactly two and the function takes the value $1/2$ irrespective of the choice of $\beta_{FBC}$; and as the frequency moves away from the band, the denominator grows without bound, and the function decays monotonically towards zero. We note that the exponent $2\beta_{FBC}$ guarantees that the function is symmetric about $\omega_c$ and remains positive for all arguments. Because the level $1/2$ is attained precisely at the band limits for every admissible $\beta_{FBC}$, it provides a natural and parameter-independent threshold separating admissible from inadmissible frequencies.

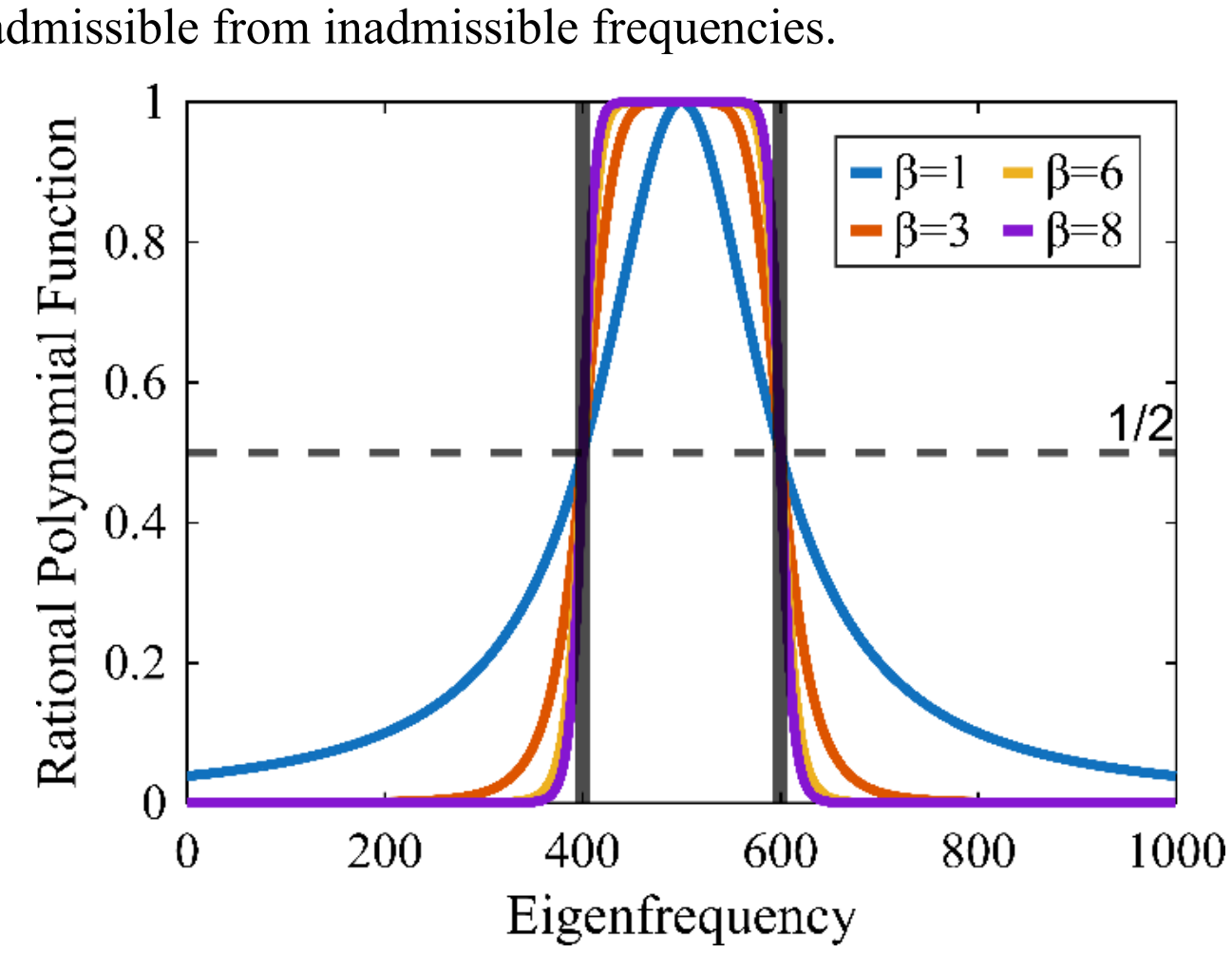

**Figure 2 Rational polynomial function:** Approximation of the Heaviside function in the frequency-band constraint.

Accordingly, the frequency band constraint (FBC) for the $b^{th}$ prohibited band is imposed in the standard non-positive form as given in ***Equation (19)***:

$$g_{FBC,b} = FBC_b - 1/2 \leq 0, \tag{19}$$

so that the satisfaction of the constraint requires $FBC_b < 1/2$, which indicates that the eigenfrequencies lie outside $[\omega_{low}^{(b)}, \omega_{upp}^{(b)}]$, as in ***Figure 2***. Unlike the Heaviside measure, $g_{FBC,b}$ varies continuously with design variables through the eigenfrequencies, and its gradient conveys both the direction and the magnitude of the change required to push a frequency out of the forbidden interval. Applying the constraint to every eigenfrequency would introduce $J_0 \times n_M$ individual constraints and would make the size of the optimization dependent on the number of modes retained. To avoid this, we gathered the contributions of all eigenfrequencies into a single scalar for each band by the p-norm aggregation as in ***Equation (20)***:

$$FBC_b = \left(\frac{1}{J_0}\sum_{j=1}^{J_0} FB_{b,j}^p\right)^{1/p}, \tag{20}$$

where $J_0$ denotes the total number of eigenvalues considered in the analysis and $p$ is the aggregation parameter. The individual contributions are obtained by evaluating the rational polynomial function at each eigenfrequency $\omega_j$ with respect to the parameters of the $b^{th}$ band, leading to an aggregated measure as:

$$FB_{b,j} = \frac{1}{1 + \left(\frac{\omega_j - \omega_c^{(b)}}{\omega_d^{(b)}}\right)^{2\beta_{FBC}}}. \tag{21}$$

The aggregated measure in ***Equation (20)*** approximates the maximum of the individual terms from below and approaches it as $p$ increases, so that a single differentiable constraint per band replaces the full set of mode-wise conditions while still being dominated by the most critical eigenfrequency. Since each term $FB_{b,j}$ is bounded between zero and unity, the aggregated quantity is likewise bounded, which keeps the constraint well scaled throughout the optimization. The two parameters $\beta_{FBC}$ and $p$ together determine how closely the smoothed constraint function reproduces the exact exclusion condition. Larger values of $\beta_{FBC}$ steepen the transition at the band edges, and thereby reduce the conservatism introduced by the smoothing; larger values of $p$ make aggregated values closer to true maximum at the cost of increased non-linearity of the constraint (see ***Section 6.2*** for detailed parameter setup).

## 4. Topology optimization for vibration attenuation

Using the proposed interpolation scheme, we introduce three topology optimization formulations for vibration attenuation. The first formulation maximizes the fundamental eigenfrequency, the second maximizes the gap between two adjacent eigenfrequencies, and the third creates a frequency gap within a targeted frequency range.

### 4.1. Maximization of fundamental frequency

The first optimization problem considered in this work aims at finding the material distribution that maximizes the fundamental eigenfrequency of the structure. Such a design is desirable in low-frequency dynamic applications, as a high fundamental frequency shifts the structural response away from low-frequency excitation. The problem is a max-min problem over the initial $J_0$ eigenfrequencies given by ***Equation (22)***:

$$\begin{aligned} \max_{\boldsymbol{x}} &: \ \min \omega_j(\boldsymbol{x}) \qquad (j = 1, \dots, J_0) \\ \text{s.t.} &: \ \frac{v_m}{v_m^*} - 1 \le 0 \qquad (m = 1, \dots, n_{Mat}) \\ &: \ 0 \le x_e^{(f)} \le 1, \quad x_e = [x_e^{(f=1)}, x_e^{(f=2)}, \dots], \end{aligned} \tag{22}$$

where $\omega_j(\boldsymbol{x})$ is the $j^{th}$ natural frequency, $v_m$ is the volume fraction of the $m^{th}$ material, and $v_m^*$ is the prescribed upper bound. The volume fraction of each material is evaluated as the element-wise average of the corresponding material weight:

$$v_m = \frac{1}{n_e} \sum_{e=1}^{n_e} w_m(e); \qquad m = 1, \dots, n_{Mat},$$

in which $w_m(e)$ denotes the weight of material $m$ in element $e$, obtained from the multi-material interpolation scheme, and $n_e$ is the total number of finite elements in the design domain. We note that ***Equation (22)*** is not suited to gradient-based optimization, as the *min* operator is non-differentiable. To overcome this difficulty, we adopt a bound formulation, in which an auxiliary scalar variable $\lambda$ is introduced that acts simultaneously as the objective function and as an additional design variable, as in ***Equation (23)***:

$$\begin{aligned} \max_{\boldsymbol{x},\lambda} &: \ \lambda \\ \text{s.t.} &: \ \frac{\lambda - \omega_j^2(\boldsymbol{x})}{\lambda_{ref}} \le 0 \qquad (j = 1, \dots, J_0) \\ &: \ \frac{v_m}{v_m^*} - 1 \le 0 \qquad (m = 1, \dots, n_{Mat}) \\ &: \ 0 \le x_e^{(f)} \le 1, \quad x_e = \left[x_e^{(f=1)}, x_e^{(f=2)}, \dots\right]. \end{aligned} \tag{23}$$

In this form, the non-smooth operator is replaced by a set of smooth inequality constraints, each of which forces one of the lowest eigenvalues to remain above the bound $\lambda$. Maximizing $\lambda$ pushes the lower part of the spectrum upwards, and at convergence, at least one of these constraints is active, so that $\lambda$ coincides with the smallest eigenvalue $\omega_1^2$ where maximizing the lowest eigenvalue is equivalent to maximizing the fundamental frequency. The quantity $\lambda_{ref}$ is a scaling factor that normalizes the eigenvalue constraints to the order of unity; it is taken as $\omega_1^2$ evaluated at the first iteration and kept fixed thereafter.

Another advantage of this formulation is that it prevents mode switching during optimization. Because all lowest modes enter the problem through separate constraints, a change in the ordering of the modes between consecutive iterations does not introduce discontinuity in the objective function. The objective remains smooth, while the reordering is preserved by constraint sets. The optimization therefore remains stable even when repeated or closely spaced eigenvalues occur, where the original max-min statement would fail. Based on the bound variable, the design vector comprises all element-wise design fields of every material together with the scalar λ:

$$\boldsymbol{x} = \left[\left\{x_e^{(1)}\right\}_{e=1}^{n_e}, \left\{x_e^{(2)}\right\}_{e=1}^{n_e}, \dots, \lambda\right].$$

### 4.2. Maximizing adjacent eigenfrequency gap

The second optimization formulation introduces a different design objective to generate a wide gap between two adjacent eigenfrequencies. Such a frequency gap is of practical interest because excitation frequencies falling inside the gap cannot be matched by any structural mode, so the design becomes effectively insensitive to a range of loading frequencies rather than to one.

$$\begin{aligned} \max_{\boldsymbol{x},\lambda_1,\lambda_2} &: \ \lambda_2 - \lambda_1 \\ \text{s.t.} &: \ \frac{\omega_j^2}{\lambda_{ref}} - \lambda_1 \le 0 \qquad (j = 1, \dots, n_b) \\ &: \ \lambda_2 - \frac{\omega_j^2}{\lambda_{ref}} \le 0 \qquad (j = n_b + 1, \dots, J_0) \\ &: \ \frac{v_m}{v_m^*} - 1 \le 0 \qquad (m = 1, \dots, n_{Mat}) \\ &: \ 0 \le x_e^{(f)} \le 1, \quad x_e = \left[x_e^{(f=1)}, x_e^{(f=2)}, \dots\right] \\ &: \ 0 \le \lambda_1 \le 1, \quad 0 \le \lambda_2 \le 1. \end{aligned} \tag{24}$$

Here, $n_b$ denotes the index of the mode below which the gap is to be created, so that the modes $j = 1, \dots, n_b$ constitute the lower eigenfrequency constraints and the modes $j = n_b + 1, \dots, J_0$ constitute upper eigenfrequency constraints. Following the same reasoning as in the previous formulation, the non-smooth statement of maximizing $\omega_{n_b+1}^2 - \omega_{n_b}^2$ is replaced by a smooth formulation in which two auxiliary scalar variables, i.e., $\lambda_1$ and $\lambda_2$, are introduced and treated simultaneously as parts of the objective and as design variables. The variable $\lambda_1$ acts as an upper bound on the normalized eigenvalues of the lower eigenfrequency constraints, since the first set of constraints forces every $\omega_j^2/\lambda_{ref}$ with $j \le n_b$ to lie below it, whereas $\lambda_2$ acts as a lower bound on the normalized eigenvalues of the upper constraint, because the second set of constraints forces every $\omega_j^2/\lambda_{ref}$ with $j > n_b$ to lie above it. Maximizing the difference $\lambda_2 - \lambda_1$ simultaneously pushes the lower constraint downwards and the upper constraint upwards.

As in the maximization of fundamental frequency, the eigenvalues are normalized by $\lambda_{ref}$, which keeps all eigenvalue constraints of the order of unity and permits the side constraints $0 \le \lambda_1 \le 1$ and $0 \le \lambda_2 \le 1$ to be imposed consistently. The remaining constraints include the volume fraction constraints, which limit the amount of each of the $n_{Mat}$ materials, and the box constraints on the element-wise design variables. Consequently, the vector of design variables contains all element-wise design variables per material together with the two bound variables:

$$\boldsymbol{x} = \left[\left\{x_e^{(1)}\right\}_{e=1}^{n_e}, \left\{x_e^{(2)}\right\}_{e=1}^{n_e}, \dots, \lambda_1, \lambda_2\right].$$

### 4.3. Maximizing targeted frequency gap

The two aforementioned formulations do not allow the frequency gap to be positioned in a targeted frequency range. In many engineering applications, however, the excitation spectrum is known in advance. Our formulation in this section introduces frequency maximization with an explicit band constraint located at a prescribed interval as in ***Equation (25)***:

$$\begin{aligned}
&\max_{\boldsymbol{x},\lambda} : \lambda \\
&\text{s.t.} : \lambda - \frac{\omega_j^2}{\lambda_{ref}} \leq 0 \quad (j = 1, \dots, J_0) \\
&\quad : \frac{v_m}{v_m^*} - 1 \leq 0 \quad (m = 1, \dots, n_{Mat}) \\
&\quad : FBC_b - \frac{1}{2} \leq 0 \quad (b = 1, \dots, n_M) \\
&\quad : 0 \leq x_e^{(f)} \leq 1,\ x_e = \left[x_e^{(f=1)}, x_e^{(f=2)}, \dots\right].
\end{aligned} \tag{25}$$

As in the maximization of fundamental frequency, the bound variable $\lambda$ serves simultaneously as the objective function and as an additional design variable. The constraints force every normalized eigenvalue $\omega_j^2/\lambda_{ref}$ of the $J_0$ modes to remain above it. Maximizing $\lambda$ therefore lifts the lower part of the spectrum, while the normalization by $\lambda_{ref}$ keeps the eigenvalue constraints of the order of unity.

In the third constraint, in which $FBC_b$ denotes the frequency-band constraint associated with the $b^{th}$ band target. This function is constructed such that it takes a value larger than one half whenever the corresponding eigenfrequency falls inside the target frequency range, and a value smaller than one half when it lies outside. Enforcing $FBC_b - 1/2 \leq 0$ for all monitored modes expels the eigenfrequencies from the prescribed interval, and creates a gap located at the targeted spectrum. The design vector for this formulation collects the element-wise design fields of individual materials together with the bound variable.

## 5. Design sensitivity derivations

To solve our optimization with a gradient-based optimizer, the sensitivities of the objective and constraint functions need to be provided at each optimization iteration [48], [49], [50]. These sensitivities are derived analytically in the following.

### 5.1. Design sensitivity of material candidates

Sensitivities are obtained by the chain rule, so the gradients of the interpolation weights with respect to design variables need to be derived. We note that the index $f$ denotes a design field, $1 \leq f \leq n_{Mat} - 1$, and $m$ denotes a material, $2 \leq m \leq n_{Mat}$, while $r$ is the penalization exponent of the SIMP interpolation. Because each un-normalized weight $w_m^{un}$ is a product of design variables raised to the power $r$, its derivative with respect to a given variable is non-zero when the variable appears in the product, which leads to three distinct cases.

#### 5.1.1. Sensitivity of the first material candidate

The unnormalized weight depends on the first design variable, and its gradient is given by ***Equation (26):***

$$\frac{\partial \boldsymbol{w}_1^{un}}{\partial \boldsymbol{\phi}_1} = r\boldsymbol{\phi}_1^{r-1} \tag{26}$$

#### 5.1.2. Sensitivity of intermediate material candidate

The gradient for intermediate materials $2 \leq m \leq n_{Mat} - 1$, depends on two cases. The first case occurs when the design field index is lower than the material index, $f < m$, the gradient follows ***Equation (27)***:

$$\frac{\partial \boldsymbol{w}_m^{un}}{\partial \boldsymbol{\phi}_f} = \left(\prod_{\substack{j=1 \\ j\neq f}}^{m-1} (1 - \boldsymbol{\phi}_j^r)\right)(-r\boldsymbol{\phi}_f^{r-1})\boldsymbol{\phi}_m^r, \tag{27}$$

and the second case is that when the design variable index coincides with the material index, i.e., $f = m$**:**

$$\frac{\partial \boldsymbol{w}_m^{un}}{\partial \boldsymbol{\phi}_f} = \left(\prod_{j=1}^{m-1} (1 - \phi_j^r)\right) r\boldsymbol{\phi}_m^{r-1}. \tag{28}$$

#### 5.1.3. Sensitivity of the last material candidate

For the last material, i.e., $m = n_{Mat}$, the gradient of the unnormalized weight in tensor form is given by ***Equation (29)***:

$$\frac{\partial \boldsymbol{w}_m^{un}}{\partial \boldsymbol{\phi}_f} = \left(\prod_{\substack{j=1 \\ j\neq f}}^{m-1} (1 - \boldsymbol{\phi}_j^r)\right)(-r\boldsymbol{\phi}_f^{r-1}). \tag{29}$$

Since the weights used in the interpolation are normalized, the normalized gradient of weights is given by ***Equation (30)***:

$$\frac{\partial \boldsymbol{w}_m}{\partial \boldsymbol{\phi}_f} = \frac{\left(\frac{\partial \boldsymbol{w}_m^{un}}{\partial \boldsymbol{\phi}_f}\right)\sum_{m=1}^{n_{Mat}} \boldsymbol{w}_m^{un} - w_m^{un}\left(\sum_{m=1}^{n_{Mat}} \frac{\partial \boldsymbol{w}_m^{un}}{\partial \boldsymbol{\phi}_f}\right)}{\left(\sum_{m=1}^{n_{Mat}} \boldsymbol{w}_m^{un}\right)^2}. \tag{30}$$

### 5.2. Sensitivity for fundamental frequency maximization

As the bound formulation introduces an auxiliary variable ($\lambda$) alongside the density design variables ($\boldsymbol{x}$), the sensitivities of the objective and constraints must be derived with respect to both sets of variables.

#### 5.2.1. Sensitivity of objective function

Our optimization is posed as a maximization, which is solved using the method of moving asymptotes (MMA). A negative sign is introduced to recast the objective in an equivalent minimization compatible with the MMA framework***:***

$$f_0 = -\lambda. \tag{31}$$

The sensitivity of the objective with respect to the design variables is given by**:**

$$\frac{\partial f_0}{\partial x_e} = 0;\ \frac{\partial f_0}{\partial \lambda} = -1. \tag{32}$$

Combined sensitivity with respect to design variables $\boldsymbol{x}$ and $\lambda$ is given by:

$$\frac{\partial f_0}{\partial \boldsymbol{x}_{mma}} = \left[\frac{\partial f_0}{\partial \boldsymbol{x}}, \frac{\partial f_0}{\partial \lambda}\right] \tag{33}$$

#### 5.2.2. Sensitivity of eigenfrequency constraints

Our design variables are modified following the sequence of $\boldsymbol{x} \to \tilde{\boldsymbol{x}} \to \boldsymbol{\phi} \to \boldsymbol{w} \to \boldsymbol{K}, \boldsymbol{M}$. We apply the chain rule for sensitivity calculation. After calculating sensitivity with respect to the projected design variables, density filtering is applied using the chain rule. For eigen mode indices $j = 1, \dots, J_0$; the eigenfrequency constraint is given by ***Equation (34):***

$$g_j = \frac{(\lambda - \omega_j^2)}{\lambda_{ref}} \leq 0. \tag{34}$$

Its sensitivity is given by***:***

$$\frac{\partial g_j}{\partial x_e} = -\frac{1}{\lambda_{ref}} \frac{\partial \omega_j^2}{\partial x_e}. \tag{35}$$

As the sensitivity follows the chain rule, it is simplified as:

$$\frac{\partial \omega_j^2}{\partial x_e^{(f)}} = \frac{\partial \omega_j^2}{\partial \phi_e^{(f)}} \frac{\partial \phi_e^{(f)}}{\partial \tilde{x}_e^{(f)}} \frac{\partial \tilde{x}_e^{(f)}}{\partial x_e^{(f)}}, \tag{36}$$

where the first term $\partial \omega_j^2 / \partial \phi_e^{(f)}$ is further simplified***:***

$$\frac{\partial \omega_j^2}{\partial \phi_e^{(f)}} = \boldsymbol{dK}_e^{(j)} \frac{\partial E_e^{eff}}{\partial \phi_e^{(f)}} + \boldsymbol{dM}_e^{(j)} \frac{\partial \rho_e^{eff}}{\partial \phi_e^{(f)}}, \tag{37}$$

where we have $\boldsymbol{dK}_e^{(j)} = \left(\boldsymbol{u}_e^{(j)}\right)^T \boldsymbol{K}_e \boldsymbol{u}_e^{(j)}$ and $\boldsymbol{dM}_e^{(j)} = -\omega_j^2 \left(\boldsymbol{u}_e^{(j)}\right)^T \boldsymbol{M}_e \boldsymbol{u}_e^{(j)}$.

Sensitivities of effective material properties are as***:***

$$\frac{\partial E_e^{eff}}{\partial \phi_e^{(f)}} = \sum_{m=1}^{n_{Mat}} E_m \frac{\partial \boldsymbol{w}_{e,m}}{\partial \phi_e^{(f)}}; \qquad \frac{\partial \rho_e^{eff}}{\partial \phi_e^{(f)}} = \sum_{m=1}^{n_{Mat}} \rho_m \frac{\partial \boldsymbol{w}_{e,m}}{\partial \phi_e^{(f)}}. \tag{38}$$

Projection derivative term $\partial \phi_e^{(f)} / \partial \tilde{x}_e^{(f)}$ of ***Equation (36)*** is obtained from ***Equation (6)***. The term $\partial \tilde{x}_e^{(f)} / \partial x_e^{(f)}$ can be calculated through density filtering as:

$$\frac{\widetilde{\partial g_j}}{\partial \boldsymbol{x}} = \boldsymbol{H} \left( \frac{\frac{\partial g_j}{\partial \boldsymbol{x}}}{\boldsymbol{H_s}} \right). \tag{39}$$

Sensitivity with respect to the auxiliary variable $\lambda$ is:

$$\frac{\partial g_j}{\partial \lambda} = \frac{1}{\lambda_{ref}}. \tag{40}$$

Final sensitivity is computed by combining ***Equation (39)*** and ***(40)*** as in ***Equation (41)***. When computing sensitivities, filtering and Heaviside projection are applied only to the density design variables, whereas neither filtering nor projection is applied to the auxiliary variables. Sensitivity for the repeated-eigenvalue case is provided in ***Section 5.5***.

$$\frac{\partial g_j}{\partial \boldsymbol{x_{mma}}} = \left[ \frac{\partial g_j}{\partial \boldsymbol{x}}, \quad \frac{\partial g_j}{\partial \lambda} \right]. \tag{41}$$

#### 5.2.3. Sensitivity of volume constraints

Sensitivity to volume constraints for each material is calculated from the material's volume fraction. For material $m = 1, \dots, n_{Mat}$, the volume constraint is given by ***Equation (42):***

$$g_{vol,m} = \frac{v_m}{v_m^{target}} - 1 \leq 0. \tag{42}$$

The element-wise sensitivity of the volume constraint is given by ***Equation (43):***

$$\frac{\partial g_{vol,m}}{\partial x_e} = \frac{1}{v_m^{target}} \frac{\partial v_m}{\partial x_e}, \tag{43}$$

which is simplified as ***Equation (44):***

$$\frac{\partial v_m}{\partial x_e} = \frac{1}{n_e} \frac{\partial w_{e,m}}{\partial x_e}. \tag{44}$$

Applying chain rules for projection and density filtering to sensitivity gives ***Equation (45):***

$$\frac{\partial w_{e,m}}{\partial x_e^{(f)}} = \sum_{f=1}^{n_{fields}} \frac{\partial w_{e,m}}{\partial \phi_e^{(f)}} \frac{\partial \phi_e^{(f)}}{\partial \tilde{x}_e^{(f)}} \frac{\partial \tilde{x}_e^{(f)}}{\partial x_e^{(f)}}. \tag{45}$$

Sensitivity of volume constraint with respect to auxiliary variable $\lambda$ is given in ***Equation (46):***

$$\frac{\partial g_{vol,m}}{\partial \lambda} = 0. \tag{46}$$

Combining sensitivity with respect to both design variables yields ***Equation (47):***

$$\frac{\partial g_{vol,m}}{\partial \boldsymbol{x_{mma}}} = \left[ \frac{\partial g_{vol,m}}{\partial \boldsymbol{x}}, \quad \frac{\partial g_{vol,m}}{\partial \lambda} \right]. \tag{47}$$

### 5.3. Sensitivity for adjacent frequency gap maximization

For this formulation, sensitivities of the objective and constraints must be derived with respect to the density design variable ($\boldsymbol{x}$) and two auxiliary variables ($\lambda_1$ and $\lambda_2$).

#### 5.3.1. Sensitivity of the objective function

The objective indicates the separation between the two auxiliary variables and is independent of the density design variables, as given by ***Equation (48):***

$$f_0 = -(\lambda_2 - \lambda_1). \tag{48}$$

The sensitivity of the objective function with respect to the respective design variables are given by ***Equation (49):***

$$\frac{\partial f_0}{\partial x_e} = 0; \ \frac{\partial f_0}{\partial \lambda_1} = 1; \ \frac{\partial f_0}{\partial \lambda_2} = -1, \tag{49}$$

where the vanishing derivative with respect to $x_e$ confirms that the material distribution is driven purely by the eigenfrequency bound constraints, whereas the two non-zero sensitivities push $\lambda_2$ upward and $\lambda_1$ downward, widening the gap. Collecting these terms in the order required by the optimizer yields the combined sensitivity:

$$\frac{\partial f_0}{\partial \boldsymbol{x_{mma}}} = \left[ \frac{\partial f_0}{\partial \boldsymbol{x}}, \frac{\partial f_0}{\partial \lambda_1}, \frac{\partial f_0}{\partial \lambda_2} \right]. \tag{50}$$

#### 5.3.2. Sensitivity of lower eigenfrequency constraints

For eigen mode indices $j = 1, \dots, n_b$, the lower frequency constraint is given by ***Equation (51)***:

$$g_j^{(1)} = \frac{\omega_j^2}{\lambda_{ref}} - \lambda_1 \leq 0, \tag{51}$$

where the normalization by the reference eigenvalue $\lambda_{ref}$ keeps all constraint values of comparable magnitude. Differentiating ***Equation (51)*** with respect to an element density gives:

$$\frac{\partial g_j^{(1)}}{\partial x_e} = \frac{1}{\lambda_{ref}} \frac{\partial \omega_j^2}{\partial x_e}. \tag{52}$$

in which the eigenvalue sensitivity $\partial\omega_j^2/\partial x_e$ is evaluated exactly as in ***Equation (36).*** Sensitivity with respect to auxiliary variables is given by:

$$\frac{\partial g_j^{(1)}}{\partial \lambda_1} = -1; \ \frac{\partial g_j^{(1)}}{\partial \lambda_2} = 0. \tag{53}$$

Final sensitivity is given by ***Equation (54)***:

$$\frac{\partial g_j^{(1)}}{\partial \boldsymbol{x_{mma}}} = \left[\frac{\partial g_j^{(1)}}{\partial \boldsymbol{x}}, \quad \frac{\partial g_j^{(1)}}{\partial \lambda_1}, \quad \frac{\partial g_j^{(1)}}{\partial \lambda_2}\right]. \tag{54}$$

### 5.3.3. Sensitivity of upper eigenfrequency constraints

For upper eigenmodes $u = n_b + 1, \dots, J_0$, constraint is given by:

$$g_u^{(2)} = \lambda_2 - \frac{\omega_u^2}{\lambda_{ref}} \leq 0. \tag{55}$$

Sensitivity of this upper bound is given by:

$$\frac{\partial g_u^{(2)}}{\partial x_e} = -\frac{1}{\lambda_{ref}} \frac{\partial \omega_u^2}{\partial x_e}, \tag{56}$$

where $\partial\omega_u^2/\partial x_e$ is the same as ***Equation (36).*** Sensitivity with respect to auxiliary variables is given by ***Equation (57):***

$$\frac{\partial g_j^{(2)}}{\partial \lambda_1} = 0; \ \frac{\partial g_j^{(2)}}{\partial \lambda_2} = 1. \tag{57}$$

Final sensitivity is given by:

$$\frac{\partial g_j^{(2)}}{\partial \boldsymbol{x_{mma}}} = \left[\frac{\partial g_j^{(2)}}{\partial \boldsymbol{x}}, \quad \frac{\partial g_j^{(2)}}{\partial \lambda_1}, \quad \frac{\partial g_j^{(2)}}{\partial \lambda_2}\right]. \tag{58}$$

### 5.3.4. Sensitivity of volume constraints

Sensitivity of the volume constraint for each material with respect to the density design variables, $\partial g_{vol,m}/\partial \boldsymbol{x}$ is the same as ***Equation (43)***. Sensitivity of volume constraints with respect to auxiliary variables $\lambda_1$ and $\lambda_2$ is given by ***Equation (59):***

$$\frac{\partial g_{vol,m}}{\partial \lambda_1} = 0; \ \frac{\partial g_{vol,m}}{\partial \lambda_2} = 0. \tag{59}$$

Combining sensitivity with respect to all design variables yields:

$$\frac{\partial g_{vol,m}}{\partial \boldsymbol{x_{mma}}} = \left[\frac{\partial g_{vol,m}}{\partial \boldsymbol{x}}, \ \frac{\partial g_{vol,m}}{\partial \lambda_1}, \frac{\partial g_{vol,m}}{\partial \lambda_2}\right]. \tag{60}$$

## 5.4. Sensitivity for targeted frequency gap

For this formulation, sensitivities of the objective function and constraints must be derived with respect to the density design variable $(\boldsymbol{x})$ and one auxiliary variable *(λ)*.

### 5.4.1. Sensitivity of the objective function

The objective function for this formulation is defined as:

$$f_0 = -\lambda. \tag{61}$$

The sensitivity of the objective function with respect to the respective design variables is given by:

$$\frac{\partial f_0}{\partial x_e} = 0; \ \ \frac{\partial f_0}{\partial \lambda} = -1. \tag{62}$$

Combined sensitivity with respect to both design variables $\boldsymbol{x}$ and $\lambda$ is given by ***Equation (63)***:

$$\frac{\partial f_0}{\partial \boldsymbol{x_{mma}}} = \left[\frac{\partial f_0}{\partial \boldsymbol{x}}, \ \ \frac{\partial f_0}{\partial \lambda}\right]. \tag{63}$$

### 5.4.2. Sensitivity of eigenfrequency constraints

For eigen mode indices $j = 1, \dots, J_0$, the eigenfrequency constraint is given by ***Equation (64)***:

$$g_j = \lambda - \frac{\omega_j^2}{\lambda_{ref}} \leq 0. \tag{64}$$

The sensitivity of this constraint is given by ***Equation (65):***

$$\frac{\partial g_j^{(1)}}{\partial x_e} = -\frac{1}{\lambda_{ref}} \frac{\partial \omega_j^2}{\partial x_e}, \tag{65}$$

where $\partial\omega_j^2/\partial x_e$ is the same as ***Equation (36).*** Sensitivity with respect to auxiliary variables is given by:

$$\frac{\partial g_j^{(1)}}{\partial \lambda} = 1. \tag{66}$$

Final sensitivity is given by:

$$\frac{\partial g_j^{(1)}}{\partial \boldsymbol{x_{mma}}} = \left[\frac{\partial g_j^{(1)}}{\partial \boldsymbol{x}}, \quad \frac{\partial g_j^{(1)}}{\partial \lambda}\right]. \tag{67}$$

The sensitivity of the volume constraint given in ***Equation (47)*** is the same for both the fundamental frequency maximization and the prescribed frequency gap formulations, as they use the same design variables.

### 5.4.3. Sensitivity for frequency band constraint

The frequency band constraint is imposed, so that no eigenfrequency of the structure is permitted inside a prescribed band. For the $b^{th}$ band, the constraint is expressed through a rational polynomial function, and is written as:

$$g_j^{(3)} = FBC_b - 1/2 \leq 0; \quad b = 1, \dots, n_M. \tag{68}$$

Sensitivity of the frequency band constraint is given by:

$$\frac{\partial g_j^{(3)}}{\partial \boldsymbol{x}} = \frac{\partial FBC_b}{\partial \boldsymbol{x}}, \tag{69}$$

where it is further simplified:

$$\frac{\partial FBC_b}{\partial x} = \frac{FBC_b^{1-p}}{J_0} \sum_{j=1}^{J_0} F\, B_{b,j}^{p-1} \frac{\partial FB_{b,j}}{\partial \omega_j} \frac{\partial \omega_j}{\partial x} \ ; \quad \frac{\partial FB_{b,j}}{\partial \omega_j} = -2 * \beta \frac{{FB_{b,j}}^2}{\omega_d} \left(\frac{\omega_j - \omega_c}{\omega_d}\right)^{2\beta-1}, \tag{70}$$

where the terms $FBC_b$ and $FB_{b,j}$ are obtained from ***Equations (20)*** and ***(21)***, respectively. The term $\partial\omega_j/\partial\mathbf{x}$ in ***Equation (70)*** is expanded as:

$$\frac{\partial\omega_j}{\partial x_e^{(f)}} = \frac{1}{2\omega_j}\frac{\partial\omega_j^2}{\partial x_e^{(f)}}, \tag{71}$$

where $\partial\omega_j^2/\partial x_e$ is the same as ***Equation (36)***. The chain rule for projection and sensitivity is similar to the maximization of fundamental frequency. Sensitivity with respect to auxiliary variables is given by:

$$\frac{\partial g_j^{(3)}}{\partial\lambda} = 0, \tag{72}$$

The final sensitivity is given by:

$$\frac{\partial g_j^{(3)}}{\partial \boldsymbol{x}_{\boldsymbol{mma}}} = \left[\frac{\partial g_j^{(3)}}{\partial \boldsymbol{x}}, \quad \frac{\partial g_j^{(3)}}{\partial\lambda}\right]. \tag{73}$$

## 5.5. Sensitivity for repeated eigenvalues

Mode tracking is generally needed when the optimizer drives multiple eigenfrequencies close. In such scenarios, the standard sensitivity derived for distinct eigenvalues loses validity, as the sensitivity of an eigenvalue would have multiple distinct values (due to the different eigenvectors), as seen in ***Equation (37)***. In fact, it is only directionally differentiable, and its derivatives can be extracted from the subspace spanned by associated eigenvectors, as discussed in prior studies [28], [32].

Let a repeated eigenvalue of multiplicity $m_c$ be denoted by $\omega_{j_1}^2 = \omega_{j_2}^2 = \cdots = \omega_{j_{m_c}}^2 = \bar{\omega}^2$, where $\bar{\omega}^2$ is the value of the coincident eigenvalues. Since exact coincidence is generally not attained numerically, two eigenvalues are treated as repeated in the present work whenever they satisfy the closeness criterion $\left|(\omega_{j+1}^2 - \omega_j^2)/\omega_j^2\right| < 10^{-4}$. This measure is preferred over an absolute one so that the same threshold remains meaningful across the entire spectrum, irrespective of the magnitude of the eigenvalues being compared.

Once a cluster of $m_c$ coincident eigenvalues has been identified, the directional derivatives are obtained by projecting elemental stiffness and mass operators onto corresponding eigenvector subspaces, where the elemental stiffness and mass perturbation matrices are assembled as:

$$\begin{aligned} \left(\mathbf{D}_{\mathbf{K}}^{(e)}\right)_{ab} &= \left(\mathbf{u}_{\mathbf{a}}^{(e)}\right)^{T}\mathbf{K}_{\mathbf{e}}\mathbf{u}_{\mathbf{b}}^{(\mathbf{e})}\ ; \\ \left(\mathbf{D}_{\mathbf{M}}^{(e)}\right)_{ab} &= \left(\mathbf{u}_{\mathbf{a}}^{(e)}\right)^{T}\mathbf{M}_{\mathbf{e}}\mathbf{u}_{\mathbf{b}}^{(e)}, \quad \text{for a, b} = 1, \ldots, \mathrm{m_c}, \end{aligned} \tag{74}$$

in which $\boldsymbol{u}_{\boldsymbol{a}}^{(\boldsymbol{e})}$ and $\boldsymbol{u}_{\boldsymbol{b}}^{(\boldsymbol{e})}$ are the restrictions of the $a^{th}$ and $b^{th}$ eigenvectors of the repeated set to the degrees of freedom of element $e$, and $\boldsymbol{K}_{\boldsymbol{e}}$ and $\boldsymbol{M}_{\boldsymbol{e}}$ are the corresponding elemental stiffness and mass matrices. Both $\boldsymbol{D}_{\boldsymbol{K}}^{(e)}$ and $\boldsymbol{D}_{\boldsymbol{M}}^{(e)}$ are symmetric matrices of order $m_c$, and they describe how a perturbation of element $e$ redistributes energy within the degenerate subspace. Combining the two elemental perturbation matrices gives the perturbation matrix associated with the repeated eigenvalue:

$$\boldsymbol{D}^{(e)} = \boldsymbol{D}_{\boldsymbol{K}}^{(e)} - \bar{\omega}^2\boldsymbol{D}_{\boldsymbol{M}}^{(e)}. \tag{75}$$

We note that we can compute the eigenvectors $\boldsymbol{p}_k$ ($k = 1, \ldots, m_c$) from the *reduced* $\boldsymbol{D}^{(\boldsymbol{e})}$ matrix that spans the subspace of repeated eigenvalues, and $\boldsymbol{p}_k$ define the directions within the degenerate subspace along which repeated eigenvalues split under an infinitesimal design perturbation. The vectors $\boldsymbol{p}_k$ restore a well-defined, unique set of directional derivatives.

The sensitivity of each eigenvalue of the cluster with respect to the projected design variable follows the chain rule through the effective material properties:

$$\frac{\partial\omega_k^2}{\partial\phi_e^{(f)}} = \frac{\partial\omega_k^2}{\partial E_e^{eff}}\frac{\partial E_e^{eff}}{\partial\phi_e^{(f)}} + \frac{\partial\omega_k^2}{\partial\rho_e^{eff}}\frac{\partial\rho_e^{eff}}{\partial\phi_e^{(f)}}, \tag{76}$$

where the stiffness and mass contributions are given by:

$$\begin{aligned} \frac{\partial\omega_k^2}{\partial E_e^{eff}} &= \boldsymbol{p}_k^T\boldsymbol{D}_{\boldsymbol{K}}^{(e)}\boldsymbol{p}_k; \\ \frac{\partial\omega_k^2}{\partial\rho_e^{eff}} &= -\bar{\omega}^2\boldsymbol{p}_k^T\boldsymbol{D}_{\boldsymbol{M}}^{(e)}\boldsymbol{p}_k, \end{aligned} \tag{77}$$

Finally, the sensitivity is propagated backwards through the projection and filtering operations to compute physical design variables as:

$$\frac{\partial\omega_k^2}{\partial x_e^{(f)}} = \boldsymbol{H}\left(\frac{\frac{\partial\omega_k^2}{\partial\phi_e^{(f)}}\frac{\partial\phi_e^{(f)}}{\partial\tilde{x}_e^{(f)}}}{\boldsymbol{H_s}}\right). \tag{78}$$

For $m_c = 1$ the perturbation matrices reduce to scalars, $\boldsymbol{p_k}$ becomes unity, and ***Equations (76)-(77)*** collapse exactly to the classical sensitivity expression for a simple eigenvalue. Treating repeated eigenvalues in this manner suppresses oscillatory behavior and mode-switching that otherwise occur when eigenfrequencies cross during optimization iterations and thereby improves convergence stability.

# 6. Numerical implementations

Our proposed multi-material topology optimization is implemented in three optimization formulations and validated by multiple numerical experiments, including maximization of the fundamental frequency, maximization of the gap between two adjacent eigenfrequencies, and the creation of frequency gaps within a targeted frequency range. The MATLAB implementation was developed with reference to the 99-line [43] and 88-line [51] topology optimization codes and univariate mapping-based code [36], with necessary modifications to accommodate the proposed multi-material interpolation scheme and eigenfrequency-based optimization formulations. All the numerical studies were implemented in MATLAB R2025b and executed on a workstation equipped with an Intel® Core™ i7-11700K CPU @ 2.50 GHz and 16 GB of RAM.

## 6.1. Material properties

We define the material properties in ***Table 1*** for the sets of three, four, five, and seven candidate materials, respectively. Young's modulus and mass density are uniformly spaced over $E \in [1, 10]$ MPa and $\rho \in [0.1, 1]$ kgm$^{-3}$, respectively, and Poisson's ratio is fixed at $\nu = 0.3$. In all four material sets, we assume that the stiffest material has the highest density, and the most compliant material has the lowest density. We note that these two extreme materials have the same stiffness-to-density ratio, i.e., $E/\rho = 10$ MPa $\mathrm{m^3kg^{-1}}$ in every example. Because materials of the extreme properties are identical across experiments, differences in optimized designs can be attributed to the number and distribution of the materials of intermediate properties which have a lower stiffness-to-density ratio than the extreme materials. We note that the properties listed in the table serve as numerical test cases and do not represent real engineering materials.

**Table 1:** Material properties of candidate materials (Poisson's ratio ν = 0.3 for all materials)

| Number of materials | Material properties | Material1 (M1) | Material2 (M2) | Material3 (M3) | Material4 (M4) | Material5 (M5) | Material6 (M6) | Material7 (M7) |
|---|---|---|---|---|---|---|---|---|
| Three materials | Modulus of Elasticity (MPa) | 10 | 5.050 | 1 | - | - | - | - |
| | Density (Kg/m$^3$) | 1 | 0.550 | 0.100 | - | - | - | - |
| Four materials | Modulus of Elasticity (MPa) | 10 | 6.700 | 3.400 | 1 | - | - | - |
| | Density (Kg/m$^3$) | 1 | 0.700 | 0.400 | 0.100 | - | - | - |
| Five materials | Modulus of Elasticity (MPa) | 10 | 7.525 | 5.050 | 2.575 | 1 | - | - |
| | Density (Kg/m$^3$) | 1 | 0.775 | 0.55 | 0.325 | 0.100 | - | - |
| Seven materials | Modulus of Elasticity (MPa) | 10 | 8.350 | 6.700 | 5.050 | 3.400 | 1.750 | 1 |
| | Density (Kg/m$^3$) | 1 | 0.850 | 0.700 | 0.550 | 0.400 | 0.250 | 0.100 |

### 6.2. Experiment parameter setup

For implementation, the following parameters are adopted to obtain well-defined optimized topologies.

- **Penalization power:** For all numerical examples, the penalization powers for both the stiffness and mass matrices are initially set to 1.0 and are subsequently updated every 25 iterations by a factor of 1.2 until reaching a maximum value of 5.0.
  If mod (loop, 25) = 0, update $r \leftarrow \min(1.2 * p, 5)$.
- **Initial design variable:** For all examples, the initial value of design variables per element is set to 0.5. For example, the initial value for the three-material example is set as $x_e$ = [0.5, 0.5] and for four-material as $x_e$ = [0.5, 0.5, 0.5].
- **Scalar variable $\lambda$:** For the pinned beam example, the scalar variable $\lambda$ in the maximization of fundamental frequency (maximization of fundamental frequency) is initialized to 5000 with bounds defined as [0, $10^6$]. In the maximization of the gap between adjacent eigenfrequencies, the scalar variables $\lambda_1$ and $\lambda_2$ are initialized with 0.4 and 0.6, respectively, with bounds [0, 1]. In frequency gap generation between the targeted range, the scalar $\lambda$ is initialized to 0.5 with bounds of [0, 1]. The scalar variable used for the square plate and C-clamp examples is described in ***Sections 7.2*** and ***7.3***, respectively
- **Filter radius:** For the pinned beam example, the filter radius in the maximization of fundamental frequency is set to 8.0 for the five-material case and 7.0 for the seven-material design case. In maximizing the gap between adjacent eigenfrequencies, the filter radius is set to 7.0 for both the three-material and four-material cases. For creating a gap between the targeted frequency range, the filter radius is chosen as 7.0 for both single and double FBC. To obtain optimized topologies, different filter radii are adopted. The filter radius used for the square plate and C-clamp examples is described in ***Sections 7.2*** and ***7.3***, respectively.
- **Projection parameter:** For the pinned beam example, the projection parameter *($\beta_f$)* is initially set to 1.0 for all three formulations. In the maximization of fundamental frequency, for the five-material design case, the projection parameter is increased by a factor of 3 to 32, then by a factor of 1.5 every 25 iterations until reaching 300, i.e.,

  If $\beta_f \leq 32$, update $\beta_f \leftarrow \min\left(3\beta_f,\ 32\right)$,

  else $\min\left(1.5\beta_f,\ 300\right)$.

  For the seven-material design case, the projection parameter is increased by a factor of 1.5 every 25 iterations, reaching a maximum of 300. For maximization of the gap between adjacent eigenfrequencies, the projection parameter is increased by a factor of 1.5 to 32, then by a factor of 3 every 25 iterations until reaching a maximum value of 300. To create a gap between the targeted range, it is increased by 1.5 times every 25 iterations until reaching a maximum value of 300. The projection parameter update schemes adopted for the square plate and C-clamp examples are described in ***Sections 7.2*** and ***7.3***, respectively.
- **FBC steepness parameter:** For all three numerical examples, to apply the frequency band constraint, the projection parameter ($\beta_{FBC}$) of the Heaviside function is increased by a factor of 1 every 25 iterations until it reaches a maximum of 32, i.e., if mod (loop, 25) = 0, update $\beta_{FBC} \leftarrow \min(\beta_{FBC} + 1, 32)$.
- **Projection threshold:** For all numerical examples, the threshold value of the projection function is taken as 0.5.

- **P-norm parameter:** For all numerical examples, the parameter in the p-norm approximation for the maximum functions is set to 40.
- **Number of eigenfrequency modes:** For all examples, the number of eigenmodes considered in the analysis is 6.
- **Volume fractions:** For maximization of both fundamental frequency and gap between adjacent eigenfrequencies, the volume fraction of the stiffest material is set to 0.3, while the remaining volume is distributed equally among the other materials. For creating gap between the targeted frequency range, the volume fractions of all materials are identical.
- **Stopping criteria:** For all examples, the optimization is terminated once the objective function shows no further change between successive iterations and the continuation parameter has reached its maximum value.

### 6.3. Optimization algorithm

We adopt MMA [52] in our topology optimization. The complete procedure is summarized in **Algorithm 1**.

**Algorithm 1:** Multi-material topology optimization for dynamic vibration attenuation

**Inputs:** number of candidate materials $n_m$; material data $\{E_m, \nu_m, \rho_m\}$, $m = 1, \dots, n_m$; mesh data; number of eigenmodes to compute $n_b$; maximum number of iterations $N_{\max}$.
**Output:** optimized design variables $\boldsymbol{x}$.
**Initialization:** $\boldsymbol{x} \leftarrow \boldsymbol{x_0}$; loop $\leftarrow 0$;
$r_{min} \leftarrow r_{min,0}$ , $\beta \leftarrow \beta_0$, $p \leftarrow p_0$ ▷ filter radius, projection and P-norm parameters
initialize MMA parameters ($\boldsymbol{x}^{\boldsymbol{loop-2}}$, $\boldsymbol{x}^{\boldsymbol{loop-1}}$, $\boldsymbol{L}$, $\boldsymbol{U}$)
while loop < $N_{\max}$ do
    $\boldsymbol{x} \leftarrow \{x_e^{(f=1)}, x_e^{(f=2)}, \dots, x_e^{(f=n_m-1)}\}$ ▷ element-wise design fields
    $\tilde{\boldsymbol{x}} \leftarrow$ Filter ($\boldsymbol{x}$, $r_{min}$)
    $\boldsymbol{\phi} \leftarrow$ Projection ($\tilde{\boldsymbol{x}}, \beta, \eta$)
    $\boldsymbol{w} \leftarrow$ weight ($\boldsymbol{\phi}$)
    $\boldsymbol{K}(\boldsymbol{x}), \boldsymbol{M}(\boldsymbol{x}) \leftarrow$ assemble ($\boldsymbol{w}$) ▷ finite element analysis
    solve $[\boldsymbol{K}(\boldsymbol{x}) - \omega_n^2 \boldsymbol{M}(\boldsymbol{x})]\, \boldsymbol{u_n} = 0, \ n = 1, \dots, n_b$
    evaluate objective and sensitivity: $f0(\boldsymbol{x}), \partial f0/\partial \boldsymbol{x}$
    evaluate constraints and sensitivities: $\boldsymbol{f}(\boldsymbol{x}), \partial \boldsymbol{f}/\partial \boldsymbol{x}$
    $\boldsymbol{x_{mma}} \leftarrow$ MMA $(\boldsymbol{x}, f0(\boldsymbol{x}), \boldsymbol{f}(\boldsymbol{x}), \partial f0/\partial \boldsymbol{x}, \partial \boldsymbol{f}/\partial \boldsymbol{x})$
    $\boldsymbol{x} \leftarrow \boldsymbol{x_{mma}}$ ▷ update design variables
    $(\beta, p, r_{min}) \leftarrow$ update $(\beta, p, r_{min})$ ▷ continuation scheme
    If loop = $N_{\max}$ & $f0^{\text{loop}-1} \approx f0^{\text{loop}}$ then
        converged ← true ▷ check convergence
    end
    loop ← loop + 1
end while
return $\boldsymbol{x}$

## 7. Numerical examples

To validate the proposed method, three numerical examples are presented, including a pinned vibrating beam, a square plate, and a C-clamp. For all examples, the candidate materials and their corresponding properties, including Young's modulus, density, and Poisson's ratio, are listed in ***Table 1***.

### 7.1. Vibrating beam

In this study, a two-dimensional beam with dimensions 8 m × 1 m is considered (see ***Figure 3***), where the midpoint of both ends of the beam is fixed. The design domain is discretized into 400×50 finite elements.

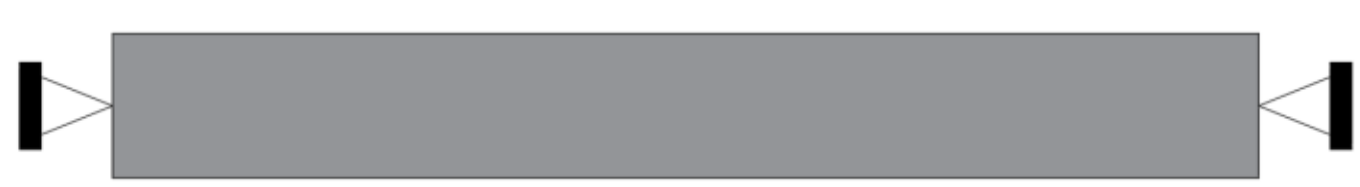

**Figure 3** Boundary condition of the pinned vibrating beam

#### 7.1.1. Maximum fundamental frequency

In this example, the objective is to maximize the fundamental eigenfrequency. The evolution of the fundamental frequency over the optimization by five material candidates is shown in ***Figure 4(c)***. The maximum fundamental eigenfrequency attained is 167 rad/s. The corresponding material distribution is presented in ***Figure 4(d)***, where color bar denotes the materials ordered from the stiffest and densest (M1) to the least stiff and dense (M5).

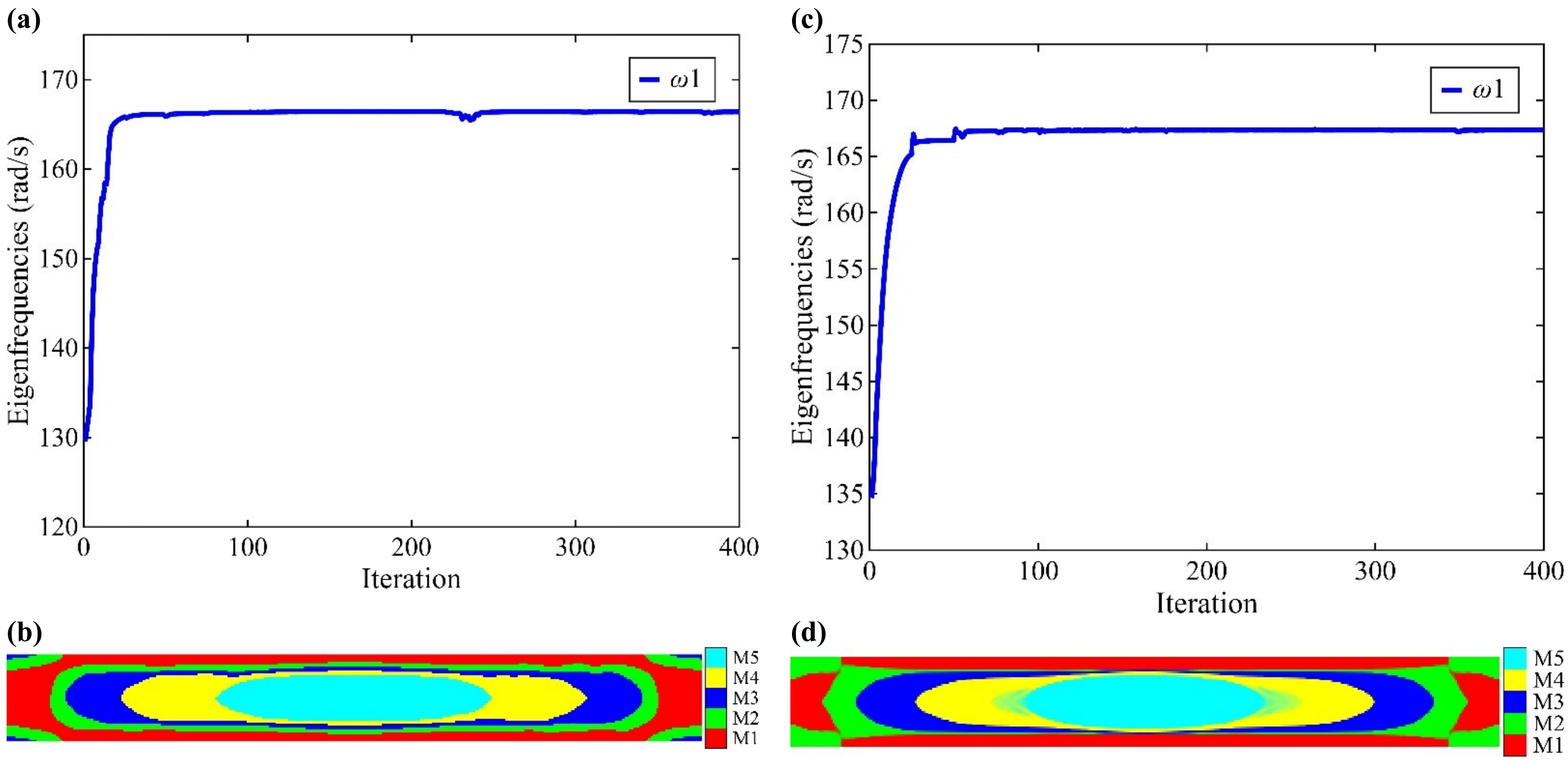

**Figure 4 Comparison of fundamental eigenfrequency maximization with five candidate materials. (a)** Convergence history of the fundamental eigenfrequency of the ordered SIMP interpolation, **(b)** optimized multi-material design, **(c)** convergence history of the fundamental eigenfrequency of the proposed multivariate weight-based interpolation, and **(d)** optimized multi-material design.

To assess the performance of the proposed framework, results are compared with those obtained using the ordered SIMP scheme [35]. The ordered SIMP interpolation was originally developed for compliance minimization problems; however, for a fair comparison, it is extended here to the eigenfrequency maximization problem. The convergence history and the optimized topologies are shown in ***Figure 4***. Both methods converge rapidly and yield similar layered material distributions (***Figure 4(b)*** and ***Figure 4(d)***), while our method achieves a slightly higher fundamental eigenfrequency of 167 rad/s, compared with 166 rad/s for the ordered SIMP method as seen in ***Figure 4(a)*** and ***Figure 4(c)***.

The beam is further optimized using seven candidate materials. As shown in ***Figure 5(a)***, the convergence is similar to that of the five-material case, with a maximum fundamental eigenfrequency of 166 rad/s. This indicates that increasing the number of candidate materials does not necessarily result in a higher fundamental eigenfrequency. The material distribution is presented in ***Figure 5(b)***.

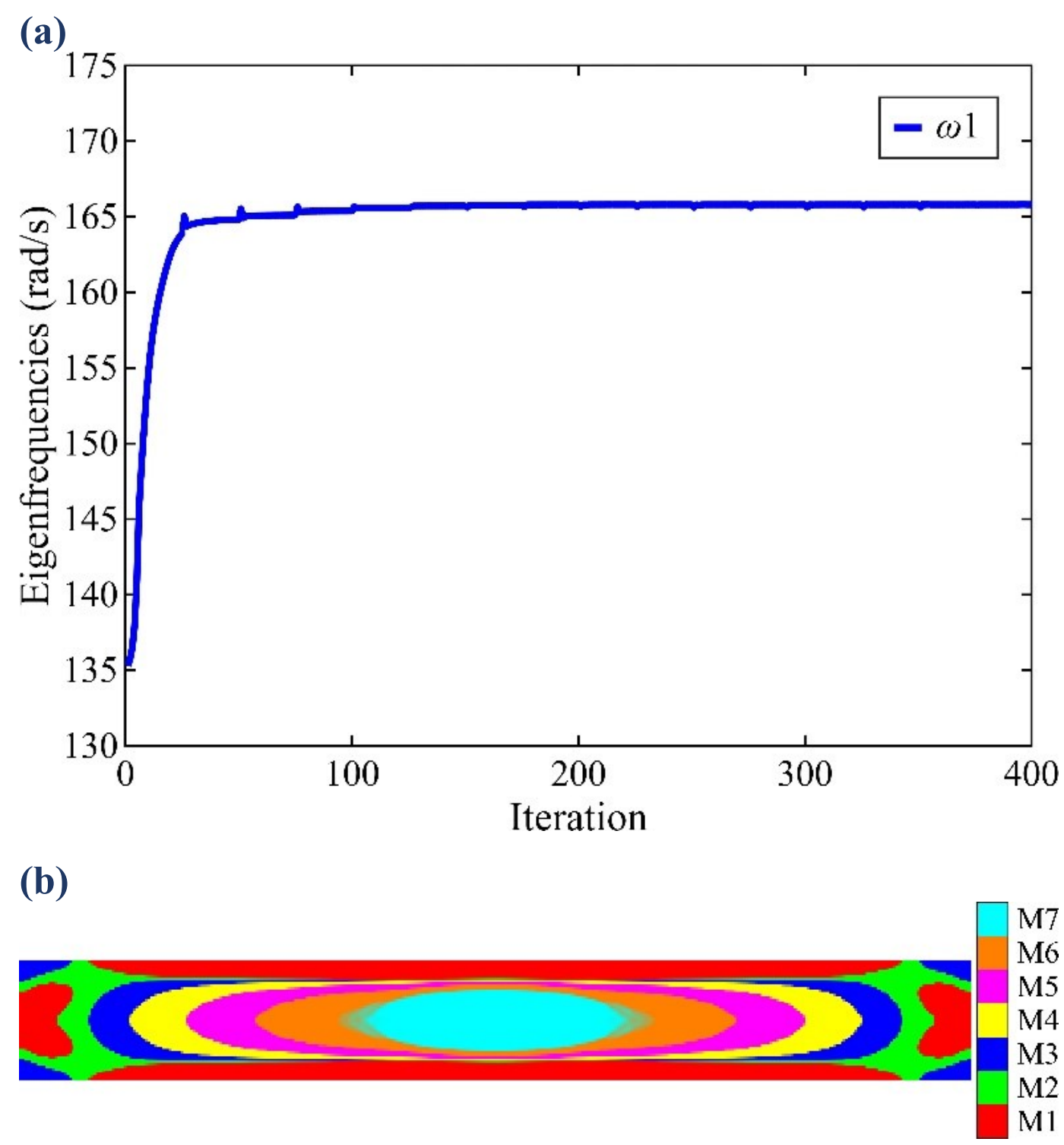

**Figure 5 Fundamental eigenfrequency maximization with seven candidate materials: (a)** convergence of the fundamental eigenfrequency and **(b)** multi-material design.

We note that the maximization of the fundamental eigenfrequency is governed by the Rayleigh quotient, i.e., $\omega^2 = (\boldsymbol{v}^{\mathrm{T}}\boldsymbol{K}\boldsymbol{v})/(\boldsymbol{v}^{\mathrm{T}}\boldsymbol{M}\boldsymbol{v})$, in which the numerator represents the modal strain energy and the denominator represents the modal kinetic energy. The eigenvalue sensitivity in ***Equation (79)*** with respect to the design variable indicates that the increase in stiffness raises the eigenvalue if the resulting gain in strain energy exceeds $\omega^2$ times the accompanying gain in kinetic energy.

$$\frac{\partial \omega^2}{\partial \boldsymbol{x}} = \boldsymbol{v}^{\mathrm{T}} \left( \frac{\partial \boldsymbol{K}}{\partial \boldsymbol{x}} - \omega^2 \frac{\partial \boldsymbol{M}}{\partial \boldsymbol{x}} \right) \boldsymbol{v}. \tag{79}$$

We find that the optimization tends to assign the stiffest and densest material to regions with high modal strain energy. Conversely, the least stiff and dense material is retained in the interior region, where reducing mass is more beneficial than increasing stiffness.

### 7.1.2. Gap maximization between adjacent eigenfrequencies

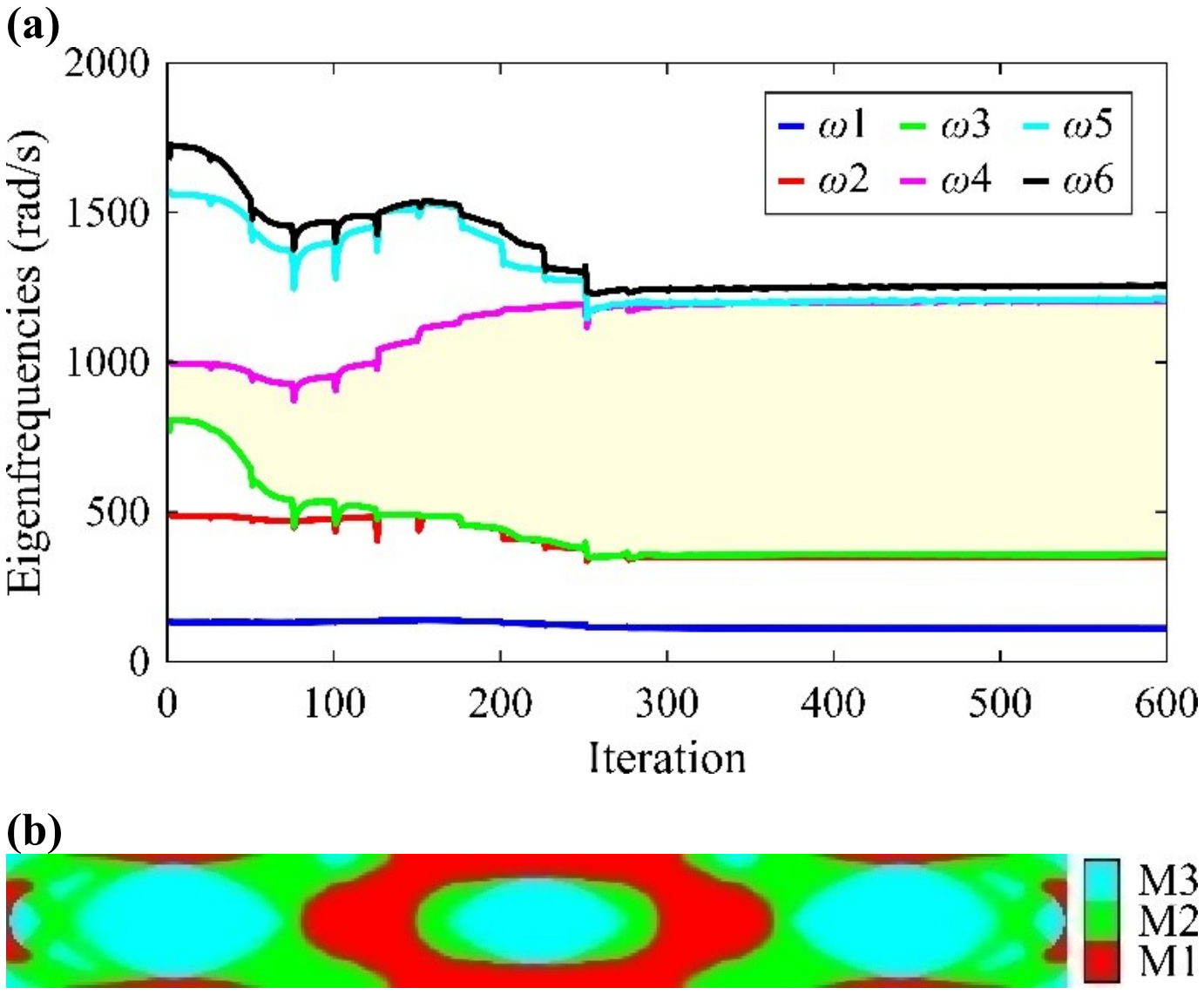

**Figure 6 Gap maximization of adjacent eigenfrequencies with three candidate materials: (a)** convergence history of the eigenfrequency gap, and **(b)** multi-material design.

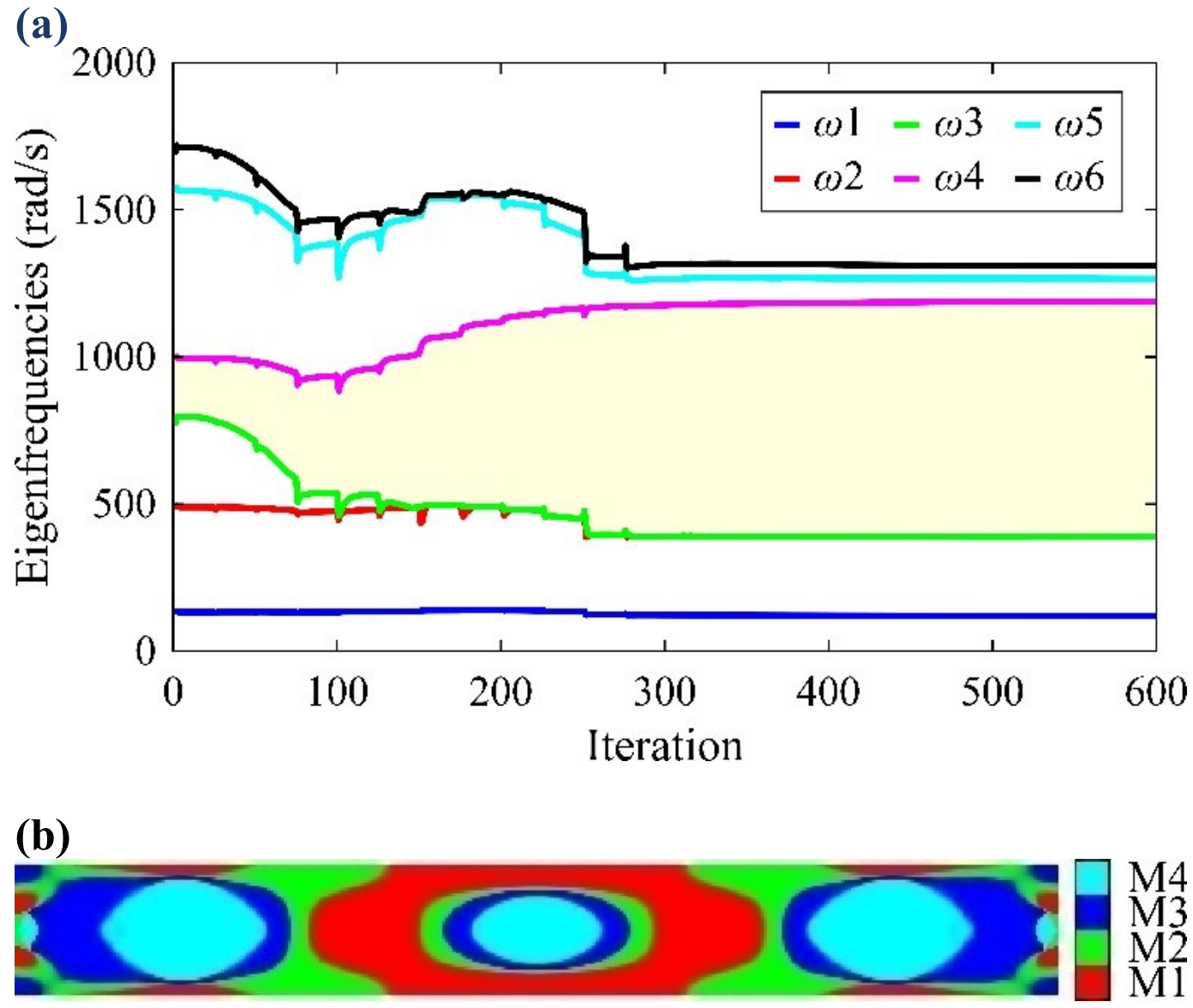

**Figure 7 Gap maximization of adjacent eigenfrequencies with four candidate materials: (a)** convergence history of the eigenfrequency gap, and **(b)** optimized multi-material design.

In this formulation, the objective is to maximize the frequency gap between the third and fourth eigenfrequencies. ***Figure 6(a)*** and ***Figure 7(a)*** present the evolution of the two bounding eigenfrequencies over the optimization history for the three- and four-material cases, respectively. In both cases, the gap widens, i.e., the third eigenfrequency moves downward while the fourth is driven upward. The optimized gaps obtained are 847 rad/s for the three-material design and 800 rad/s for the four-material design, as shown in ***Figure 6(b)*** and ***Figure 7(b)***. The topologies differ from those obtained in the preceding formulation. In particular, the stiffest and densest material is no longer concentrated in the regions adjacent to the support regions. The difference stems from the objective, which is governed not by a single eigenmode, but by the relative positions of two adjacent modes. Material layout is dictated by the difference between the modal energy distributions of the third and fourth modes.

We show the first six mode shapes of the four-material design in ***Figure 8,*** where colorbars represent mass-normalized displacement magnitudes. We note that modes $\omega_1$, $\omega_2$, $\omega_4$, $\omega_5$ and $\omega_6$ are transverse bending modes, and mode $\omega_3$ is a longitudinal (axial) mode, in which the beam stays essentially straight and moves along its axis. Unlike the global bending modes $\omega_1$, $\omega_2$ and $\omega_4$, modes $\omega_5$ and $\omega_6$ are localized near the supports, where the end segments deflect significantly.

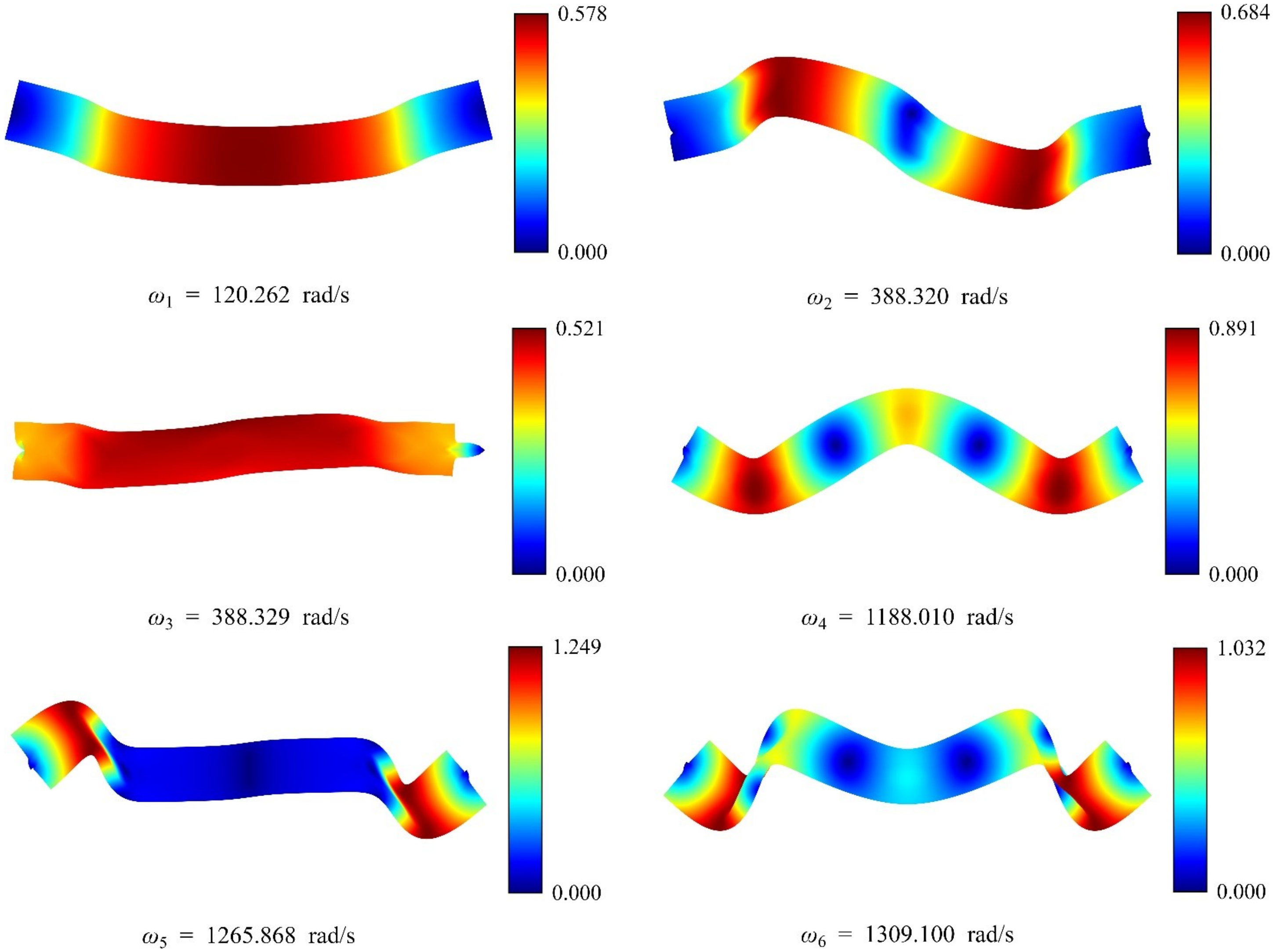

**Figure 8 Vibrational eigenmodes:** First six eigenmodes of the four-material pinned beam obtained from eigenfrequency-gap maximization in ***Section 7.1.2.***

### 7.1.3. Creating a gap between targeted frequency range

Unlike the preceding formulations, in which the target eigenfrequencies are identified by their modal index, this formulation imposes a frequency-band constraint that prevents any natural frequency of the structure from lying within a user-specified operating range, thereby creating a gap over that range. The proposed rational polynomial function is employed to approximate the Heaviside function to avoid non-differentiability in the FBC formulation for gradient-based optimization. Two cases are considered. The first involves a single operating range of $[500, 1000]$ rad/s, and the second considers two different operating ranges of $[300, 500]$ rad/s and $[700, 1000]$ rad/s. In both cases, the structure must be designed such that none of its natural frequencies lies within the prescribed intervals.

#### 7.1.3.1. Design for a single frequency range

We start by designing with three candidate materials, such that all natural frequencies are outside the prescribed frequency interval of $[500, 1000]$ rad/s. The resulting eigenfrequencies are reported in ***Table 2***, the convergence history in ***Figure 9(a)***, and the corresponding material distribution in ***Figure 9(b)***, respectively. The attained gap slightly exceeds the prescribed interval, which is attributable to the approximation error of the p-norm aggregation.

**Table 2:** First six eigenfrequencies ($\omega_i$) of the three-material beam by excluding natural frequencies from a single targeted frequency range of $[500, 1000]$ rad/s and from two targeted frequency ranges of $[300, 500]$ and $[700, 1000]$ rad/s

| Eigenmode index ($i$) | 1 | 2 | 3 | 4 | 5 | 6 |
|---|---|---|---|---|---|---|
| Single FBC, $\omega_i$ (rad/s) | 133 | 335 | 1000 | 1001 | 1086 | 1386 |
| Double FBC, $\omega_i$ (rad/s) | 133 | 258 | 605 | 629 | 700 | 1011 |

#### 7.1.3.2. Design for two frequency ranges

In this case, the structure is designed to create frequency gaps over two disjoint operating ranges, $[300, 500]$ rad/s and $[700, 1000]$ rad/s. The frequency-band constraint is accordingly extended to act on both intervals, so that no eigenfrequency is admitted into either gap. The resulting eigenfrequencies for the three-material design are reported in ***Table 2***, the convergence in ***Figure 10(a)***, and material distributions in ***Figure 10(b)***, respectively.

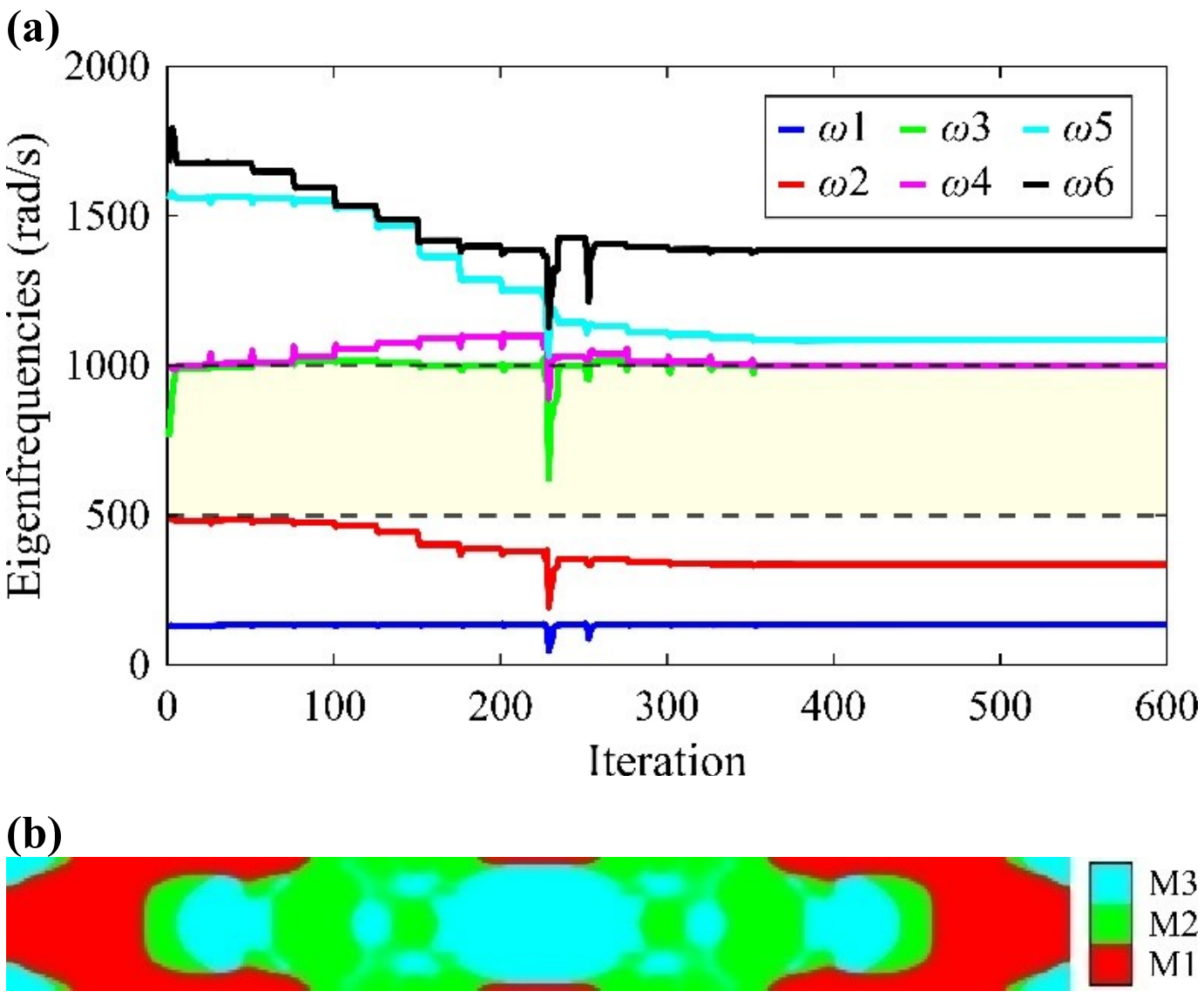

**Figure 9 Targeted frequency gap for a three-material beam: (a)** convergence of the frequency gap and **(b)** optimized multi-material design.

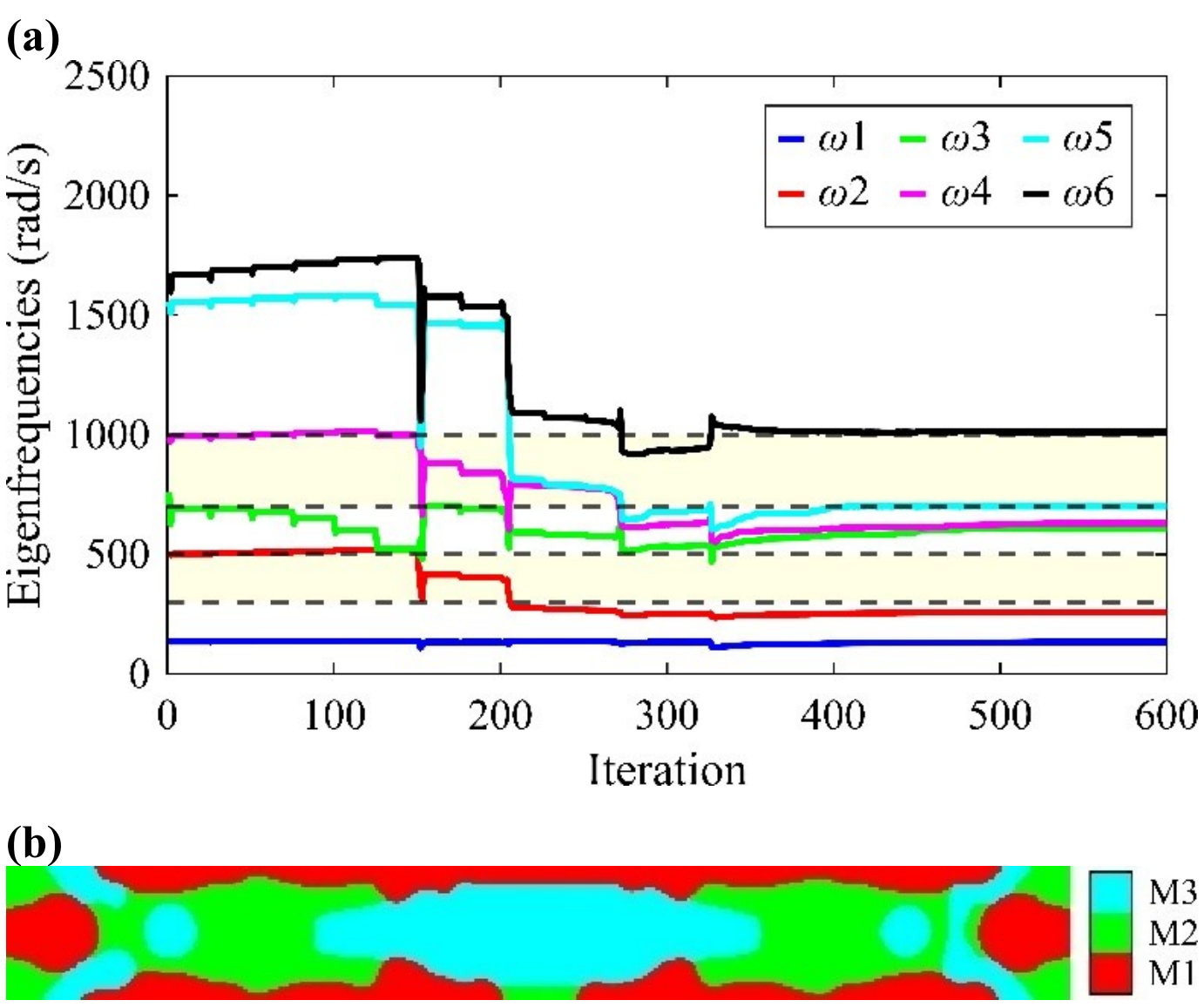

**Figure 10 Two targeted frequency gaps: (a)** convergence of the frequency gap, and **(b)** optimized multi-material design.

The convergence history shows minor variation in the eigenfrequencies during the initial iterations, followed by a sharp drop in eigenfrequencies at the 150$^{th}$ iteration, caused by the update of the continuation parameters; thereafter, the spectrum stabilizes rapidly and satisfies the frequency-band constraints for both intervals. It is noted that the transition occurs earlier than in a single frequency band, which is consistent with the additional constraint acting to drive the design away from the initial configuration at an earlier stage of the optimization.

### 7.2. Square plate design

A two-dimensional square plate with dimensions of 1 m × 1 m is shown in ***Figure 11***, with all four corner points fixed. The design domain is discretized using 200×200 finite elements. The continuation for the projection parameter $\beta_f$ is adjusted to obtain optimized topologies with minimal intermediate-density regions. The initial value of $\beta_f$ is set to 2 for the maximization of fundamental frequency and 1 for maximizing the gap between adjacent eigenfrequencies and creating a gap between the targeted eigenfrequency range. In the maximization of fundamental frequency, $\beta_f$ is increased by a factor of 3 until it reaches 32, and thereafter by a factor of 1.5 every 25 iterations up to a maximum value of 300. In maximizing gap between adjacent eigenfrequencies, $\beta_f$ is increased by a factor of 1.5 every 25 iterations until it reaches 300. In creating a gap between the targeted frequency range, $\beta_f$ is increased by a factor of 1.5 until it reaches 32, and then by a factor of 3 every 25 iterations up to 300. In addition, the filter radius is set to 8 for all three formulations. The scalar variable $\lambda$ in the maximization of fundamental frequency is initialized to 5000, with bounds [0, $10^8$]. In maximizing the gap between adjacent eigenfrequencies, scalar variables $\lambda_1$ and $\lambda_2$ are initialized with 0.4 and 0.6, respectively, with bounds [0, 1]. To create a gap between the targeted range, the variable $\lambda$ is initialized to 0.5 with bounds of [0, 1].

#### 7.2.1. Maximum fundamental frequency

In this numerical example, the fundamental eigenfrequency of a two-dimensional square plate is maximized, and the optimized material distribution for the five-material case is presented in ***Figure 12(b)***. As in the pinned beam example, the stiffest and densest material concentrates near the supports. The convergence history is shown in ***Figure 12(a)***.

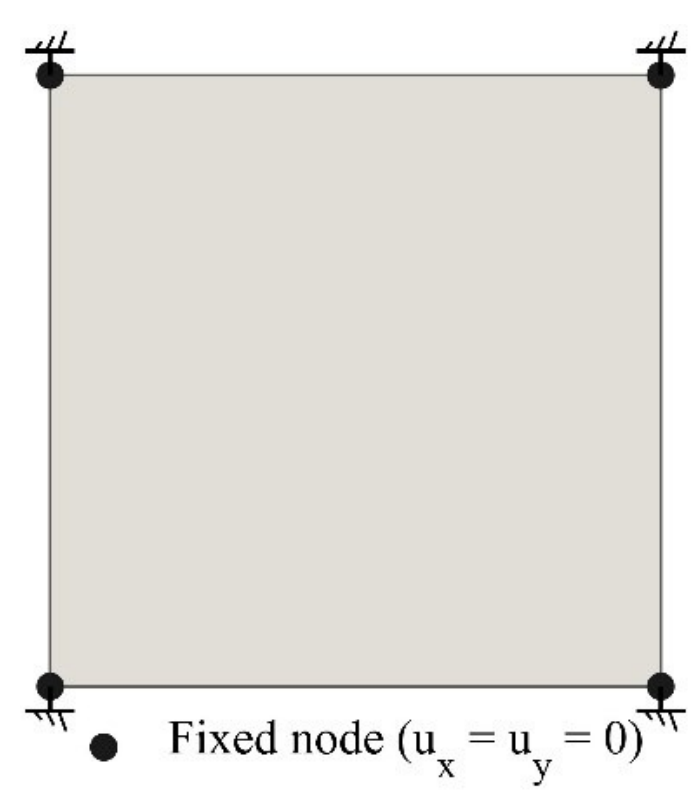

**Figure 11 Square plate:** boundary condition and design domain.

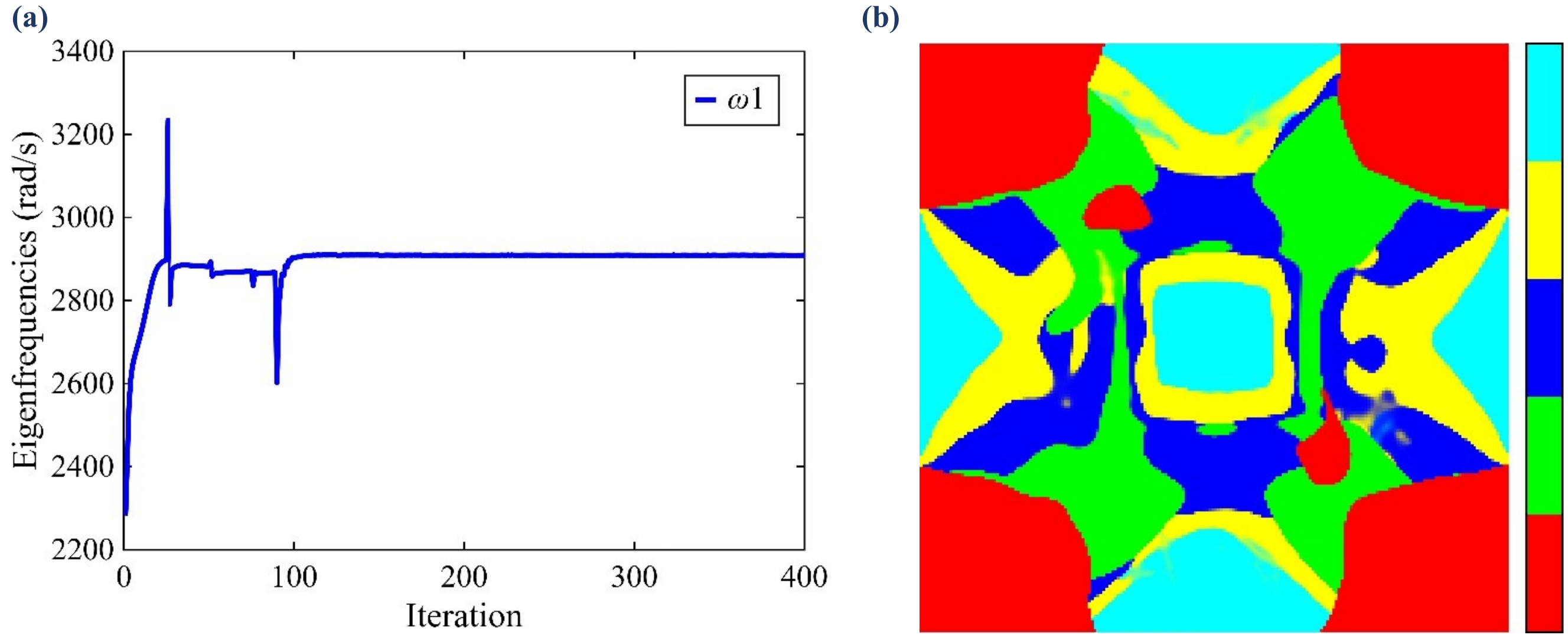

**Figure 12 Fundamental eigenfrequency maximization with five candidate materials: (a)** convergence history of the fundamental eigenfrequency and **(b)** optimized multi-material design.

### 7.2.2. Maximum gap between adjacent eigenfrequencies

In this numerical example, gap maximization between adjacent eigenfrequencies is applied to the square plate to maximize the gap between the third and fourth eigenfrequencies, and a maximized frequency gap of 8,499 rad/s is obtained for the five-material case. The convergence history in ***Figure 13(a)*** shows that the gap is opened through a reduction of the lower eigenfrequencies. The first three eigenfrequencies decrease gradually up to approximately the 150$^{th}$ iteration, undergo a pronounced drop thereafter, and subsequently stabilize at values close to the lower bound of the spectrum, whereas the fourth and higher eigenfrequencies vary only slightly throughout the optimization. The material distribution is presented in ***Figure 13(b)***, in which the candidate materials are placed about the center of the domain.

### 7.2.3. Creating the gap between targeted frequency range

In this study, a single operating frequency range of $[2000, 10000]$ rad/s is considered for the square plate. The resulting eigenfrequencies for the five-material design are reported in ***Table 3***, and the convergence history is shown in ***Figure 14(a)***, from which a clear frequency gap is observed. The corresponding optimized material distribution is presented in ***Figure 14(b)***, where materials form concentric layers ordered from the stiffest material at the center to the least dense one at the boundary.

**Table 3:** First six eigenfrequencies ($\omega_i$) of five-material square plate design by excluding natural frequencies from a single targeted frequency range of $[2000, 10000]$ rad/s

| Mode ($i$) | 1 | 2 | 3 | 4 | 5 | 6 |
|---|---|---|---|---|---|---|
| $\omega_i$ (rad/s) | 1804 | 1804.5 | 1805 | 9992 | 9994 | 10799 |

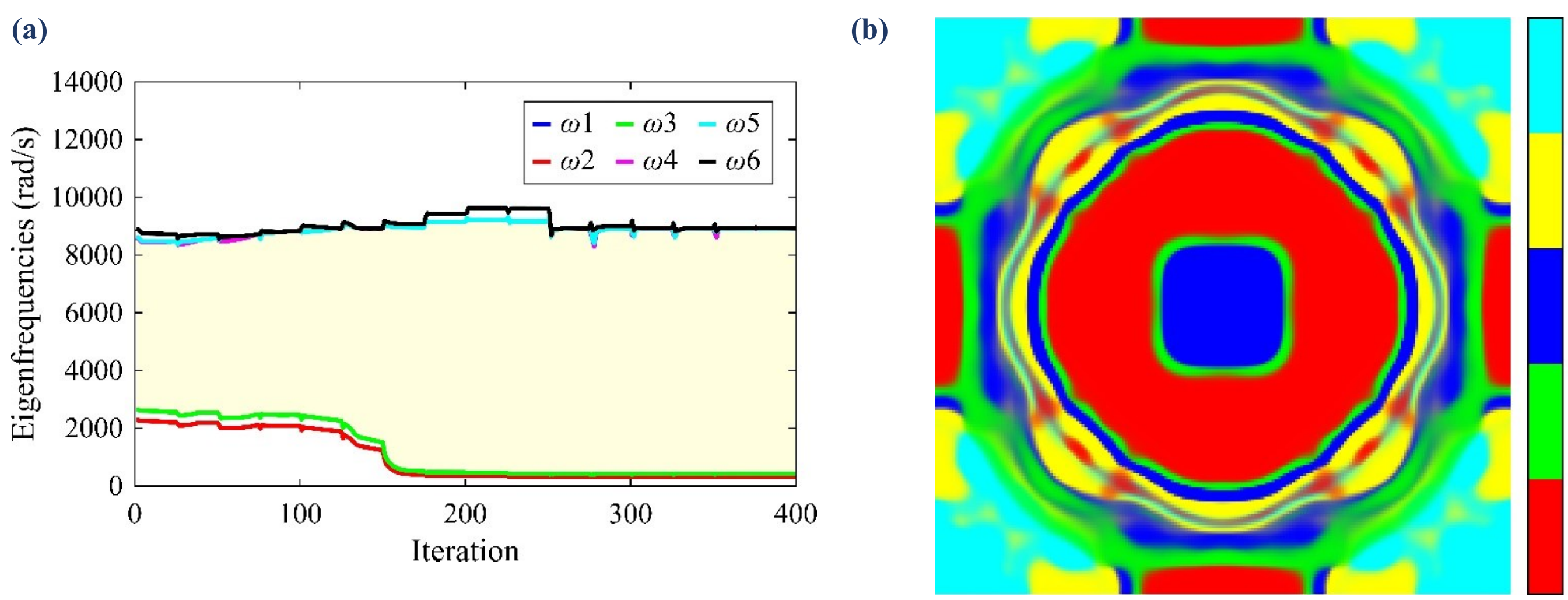

**Figure 13 Maximizing the gap between adjacent eigenfrequencies with five candidate materials: (a)** convergence history of the eigenfrequency gap, and **(b)** optimized multi-material design.

### 7.3. C-clamp design

We design a C-clamp of non-rectangular geometry in this section, as shown in ***Figure 15***, in which the right arms are fixed. The design domain is discretized using finite elements of size 5 mm × 5 mm. The three optimization formulations are considered. As in the previous examples, a continuation strategy is applied to the projection parameter $\beta_f$ to obtain optimized topologies with minimal intermediate-density regions. The initial value of $\beta_f$ is set to 1 for all three formulations. In maximization of both fundamental frequency and gap of adjacent eigenfrequencies, $\beta_f$ is increased by a factor of 3 until it reaches 32, and thereafter by a factor of 1.5 every 25 iterations up to a maximum value of 300. In creating a gap between the targeted frequency range, $\beta_f$ is increased by a factor of 1.5 until it reaches 32, and then by a factor of 3 every 25 iterations up to 300. In addition, the filter radius is set to 8 for all three formulations. The variable $\lambda$ in the maximization of fundamental frequency is initialized to 5000, with bounds $[0, 10^8]$. In maximizing the gap between adjacent eigenfrequencies, the scalar variables $\lambda_1$ and $\lambda_2$ are initialized with 0.4 and 0.6, respectively, with bounds [0, 1]. To create a gap between the targeted frequency range, the variable $\lambda$ is initialized to 0.5 with bounds of [0, 1].

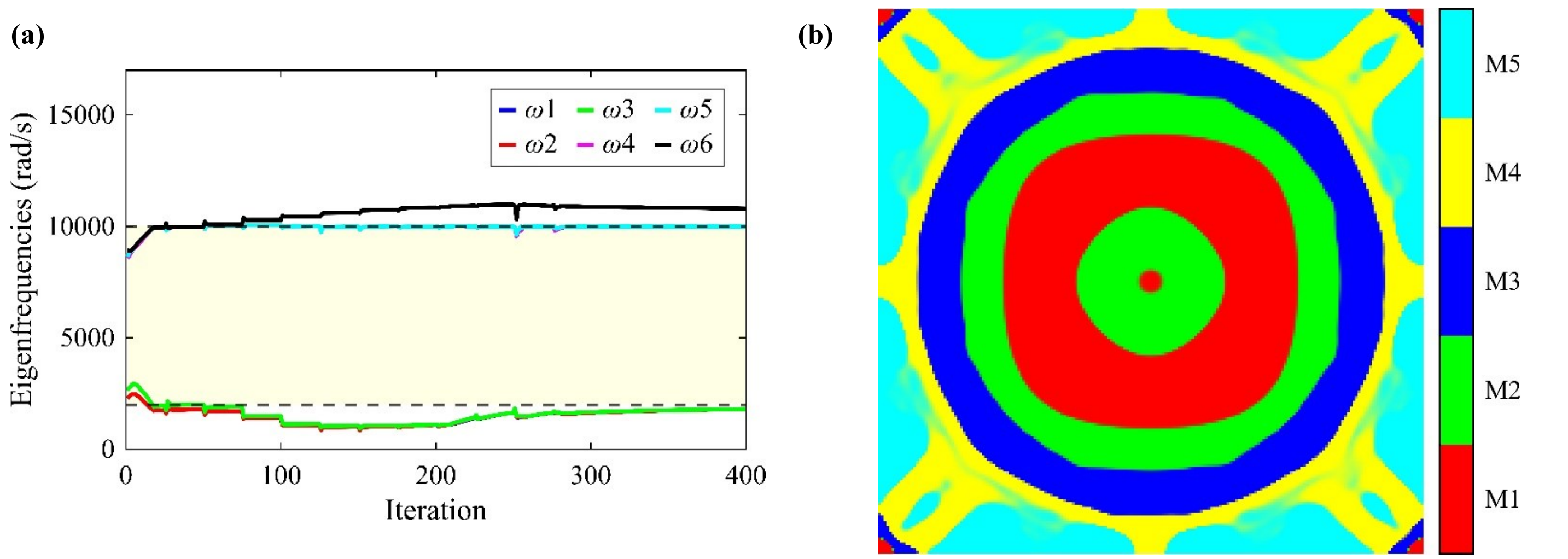

**Figure 14 Targeted frequency gap with five candidate materials: (a)** convergence history of the frequency gap and **(b)** optimized multi-material design.

#### 7.3.1. Maximum fundamental frequency

The fundamental eigenfrequency is maximized for a 2D C-clamp, and the optimized material distribution for the five-material case is presented in ***Figure 16(b)***. The convergence history is shown in ***Figure 16(a)***.

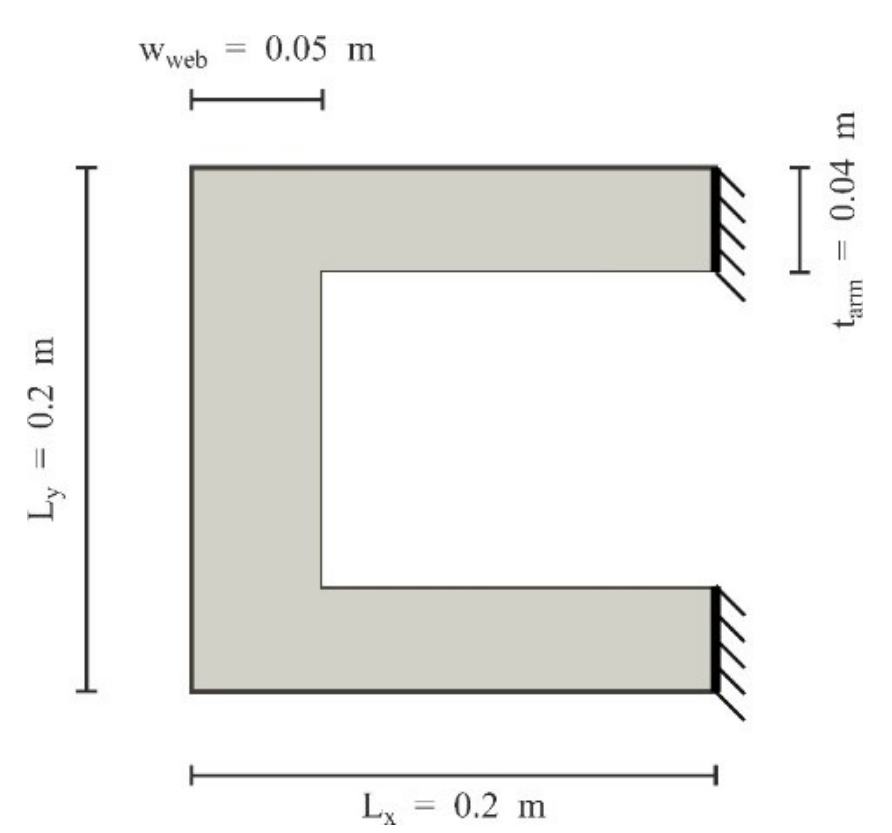

**Figure 15** Boundary condition and geometry of C-clamp.

#### 7.3.2. Maximum gap between adjacent eigenfrequencies

In this numerical example, gap maximization between adjacent eigenfrequencies is applied to the C-clamp in order to maximize the gap between the third and fourth eigenfrequencies, and a maximized frequency gap of 16285 rad/s is obtained for the five-material case. The convergence history in ***Figure 17(a)*** indicates that the gap is opened predominantly through a reduction of the third eigenfrequency rather than an elevation of the fourth. The optimized material distribution is presented in ***Figure 17(b)***. The layout is symmetric about the horizontal mid-plane. Unlike the maximization of the first eigenfrequency, the objective is to increase the fourth eigenfrequency while decreasing the third, widening the frequency gap between them.

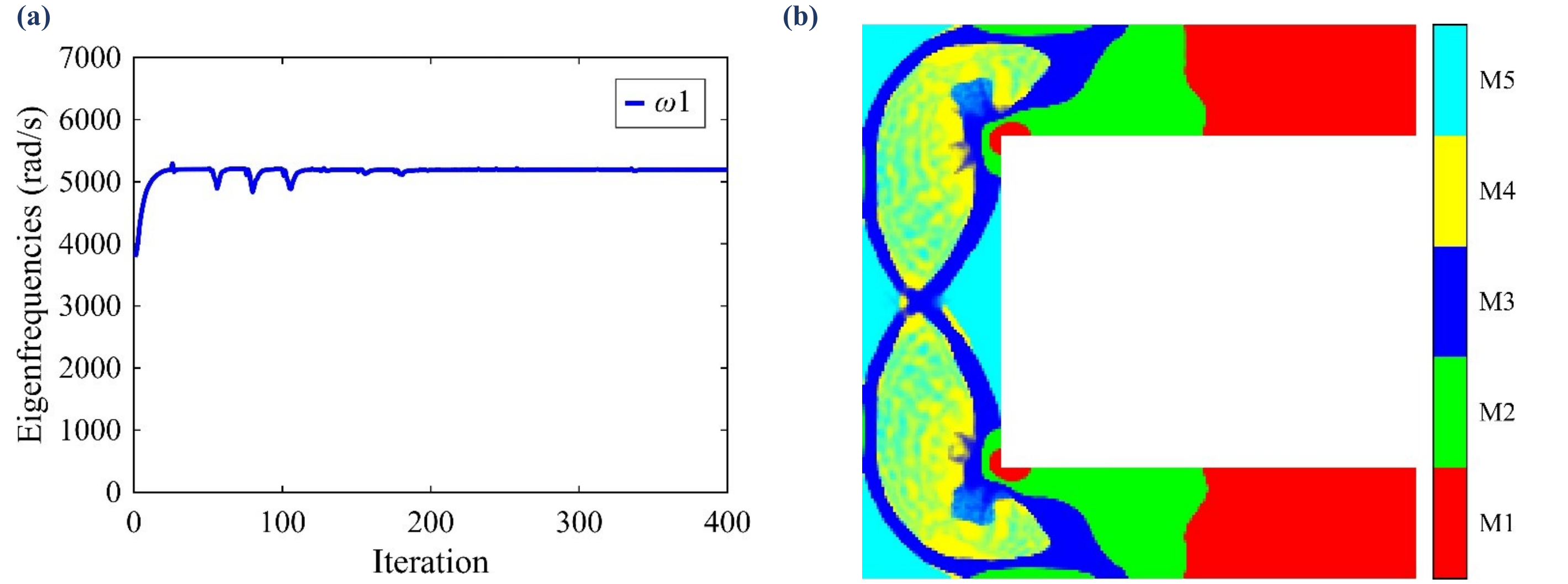

**Figure 16 Fundamental eigenfrequency maximization with five candidate materials: (a)** convergence history of the fundamental eigenfrequency and **(b)** optimized multi-material design.

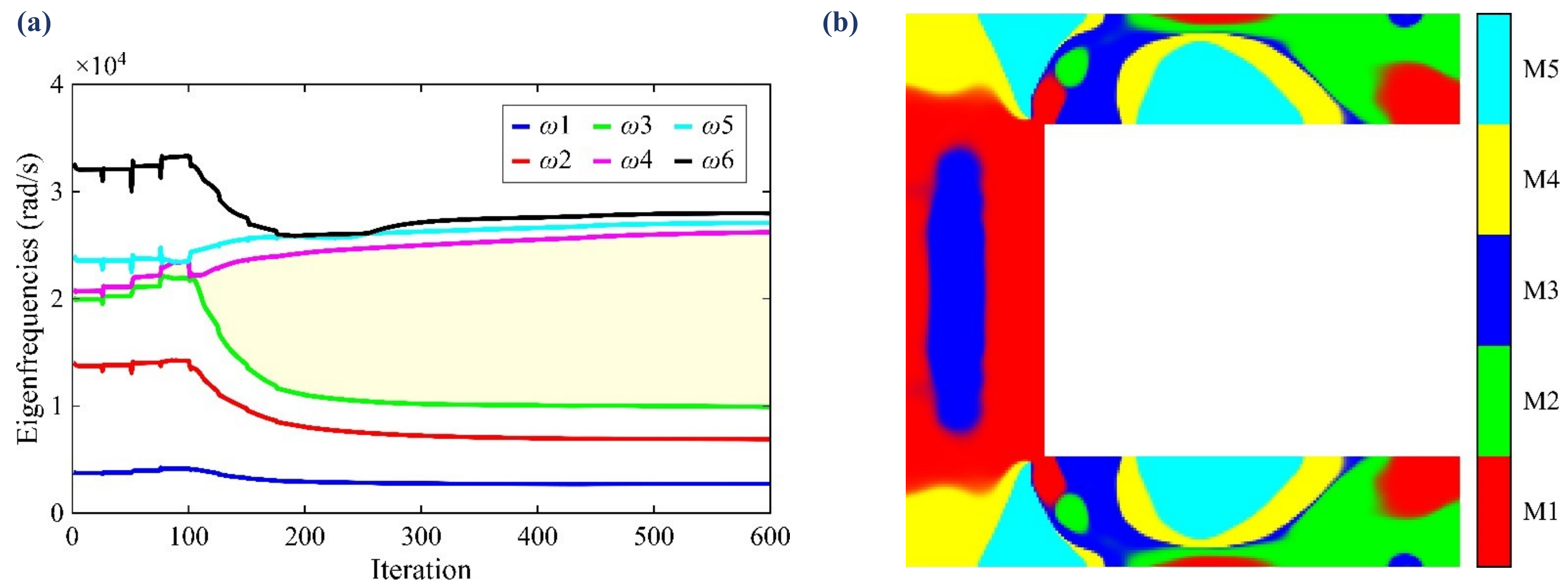

**Figure 17 Maximizing the gap between adjacent eigenfrequencies with five candidate materials: (a)** convergence history of the eigenfrequency gap and **(b)** optimized multi-material design.

### 7.3.3. Creating gaps between targeted frequency ranges

A single frequency gap of $[10000, 25000]$ rad/s is considered. The resulting eigenfrequencies for the five-material configuration are reported in ***Table 4,*** and the convergence history is shown in ***Figure 18(a)***, from which a clear frequency gap spanning the interval $[10000, 25000]$ rad/s is observed. The optimized material distribution is presented in ***Figure 18(b)***.

**Table 4:** First six eigenfrequencies ($\omega_i$) of five-material C-clamp obtained by excluding natural frequencies from a single targeted frequency range of [10000, 25000] rad/s

| Mode ($i$) | 1 | 2 | 3 | 4 | 5 | 6 |
|---|---|---|---|---|---|---|
| $\omega_i$ (rad/s) | 4124 | 10332 | 24472 | 24725 | 25992 | 30637 |

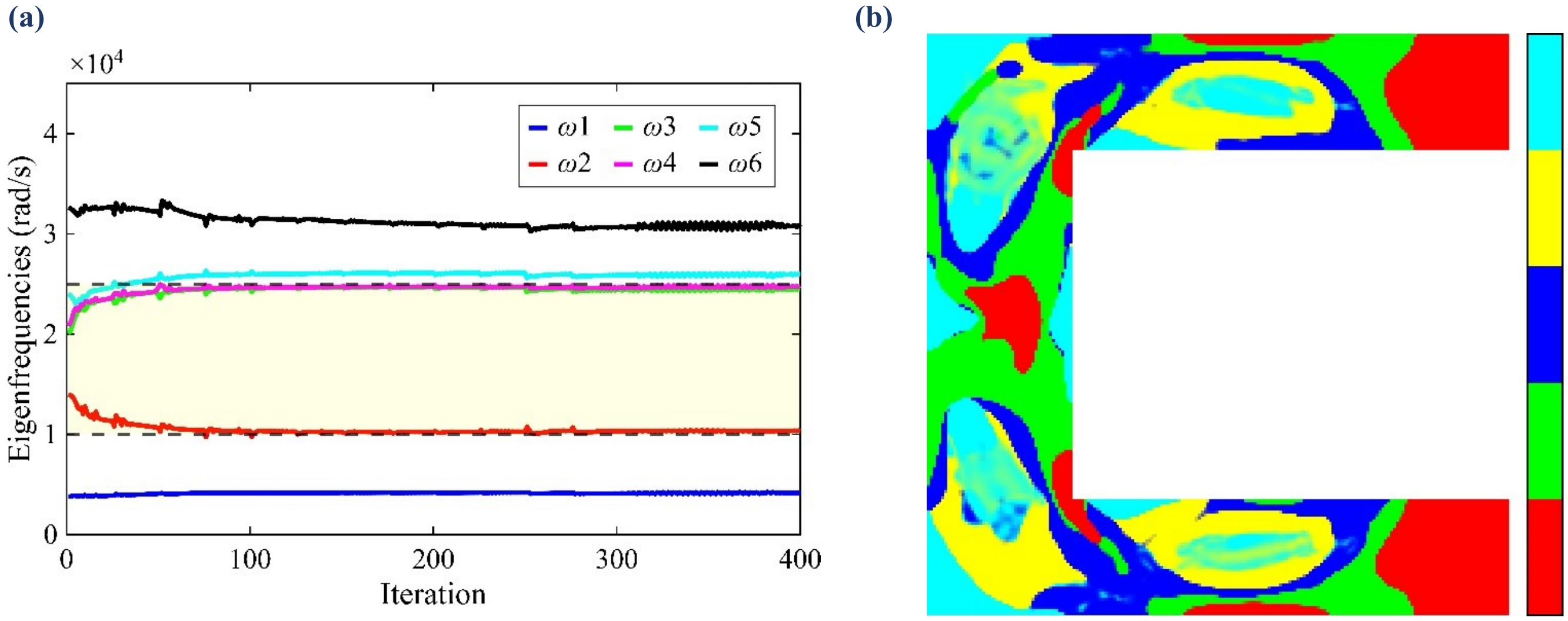

**Figure 18 Targeted frequency gap with five candidate materials: (a)** convergence history of the frequency gap and **(b)** optimized multi-material design.

## 8. Conclusions

This study presented a novel MMTO framework with multivariate weight-based interpolation for structural vibration attenuation. Specifically, we developed three optimization formulations for multi-material designs: maximization of the fundamental eigenfrequency, maximization of the frequency gap between adjacent eigenmodes, and exclusion of natural frequencies from a prescribed interval, where we adopted an auxiliary scalar design variable within a bound formulation to replace direct minimum and maximum objective functions with individual eigenvalue constraints. We observed that the three formulations produced distinct structural dynamic responses in numerical experiments. By fundamental-eigenfrequency maximization, we increase the lowest natural frequency and thereby delay the onset of the first structural resonance. Through adjacent-frequency-gap maximization, we tailored material distribution to separate targeted neighboring eigenfrequencies and enlarge their interval. Through frequency-band constraint, natural frequencies were shifted outside a prescribed operating interval, thereby avoiding structural resonances, where we proposed a new rational polynomial function for smooth approximation of discontinuous frequency-band constraints to enable sensitivity evaluation.

Future work will extend the proposed framework toward programmable vibration attenuation, where structural dynamic characteristics can be adjusted in response to external stimuli. It may also be extended to the design of multi-material metamaterials and periodic structures for vibration and wave attenuation. Further developments toward multiphysics applications, including coupled mechanical, thermal, acoustic, or electromechanical systems, would broaden the applicability of the proposed approach to more complex engineering systems.

## Acknowledgements

The authors would also like to thank the constructive suggestions from anonymous reviewers. The authors would like to thank discussions with Dr. Behzad Rankouhi from the University of Minnesota, Dr. Cheng Huang from the University of Kansas, and Dr. Nathan Hertlein from Air Force Research Laboratory. The authors would like to thank Prof. Krister Svanberg for the MATLAB codes for the MMA optimizer. Shiguang Deng would like to thank the support of the Faculty Startup Grant, the New Faculty Research Development Grant, and the General Research Grant from the University of Kansas.

## Replication of results

The results presented should be reproducible through the implementation of the methodologies described in this publication.

## Ethics approval and Consent to participate

Not applicable.

## Data availability

All data involved are included in the paper.

## Conflict of interest

No competing financial or non-financial interests.